\documentclass[notitlepage,a4paper,aps,PRD,preprintnumbers,nofootinbib,showkeys,superscriptaddress,11pt]{revtex4-1}

\pdfoutput=1
\usepackage[normalem]{ulem}
\usepackage{fullpage}
\usepackage{amsfonts}
\usepackage{amsmath}
\usepackage{slashed}
\usepackage{amssymb}
\usepackage{graphicx}
\usepackage{makeidx}
\usepackage{cancel}
\usepackage{epic}
\usepackage{eepic}
\usepackage{epsfig}
\usepackage{latexsym}
\usepackage[dvipsnames]{xcolor}
\usepackage{float}
\usepackage{multirow}
\usepackage[export]{adjustbox}
\usepackage{xurl,hyperref}
\usepackage{enumitem}
\hypersetup{colorlinks=true,citecolor=red,linkcolor=NavyBlue,urlcolor=NavyBlue}
\usepackage[utf8]{inputenc}
\usepackage[T1]{fontenc}
\usepackage[caption=false]{subfig}
 
\usepackage{natbib}
\usepackage{relsize}
\usepackage[left=2.5cm,right=2.5cm,top=2.1cm,bottom=2.2cm]{geometry}
\usepackage{mathptmx}
\begin{document}
\relscale{1.0}
\captionsetup[subfigure]{labelformat=empty}

\title{~\\Cancellation in Dark Matter-Nucleon Interactions: the Role of 
Non-Standard-Model-like Yukawa Couplings }

\author{Debottam Das}
\email{debottam@iopb.res.in}
\affiliation{Institute of Physics, Sachivalaya Marg, Bhubaneswar 751 005, India}
\affiliation{Homi Bhabha National Institute, Training School Complex, Anushakti Nagar, Mumbai 400 094, India}

\author{Bibhabasu De}
\email{bibhabasu.d@iopb.res.in}
\affiliation{Institute of Physics, Sachivalaya Marg, Bhubaneswar 751 005, India}
\affiliation{Homi Bhabha National Institute, Training School Complex, Anushakti Nagar, Mumbai 400 094, India}

\author{Subhadip Mitra}
\email{subhadip.mitra@iiit.ac.in}
\affiliation{Center for Computational Natural Sciences and Bioinformatics, International Institute of Information Technology, Hyderabad 500 032, India}

\date{\today}
\preprint{IP-BBSR/2020-6}

\begin{abstract}
Extensive searches to probe the particle nature of dark matter  (DM) have been going on for some decades now but, so far, no conclusive evidence has been found. Among various options, the Weakly Interacting Massive Particles (WIMP) remains one of the prime 
possibilities as candidates for DM near the TeV scale. Taking a phenomenological view, such null results may be explained for a generic WIMP in a Higgs-portal scenario if we allow the light-quark Yukawa couplings to assume non-Standard Model (non-SM)-like values. This follows from a cancellation among  different terms in the DM-nucleon scattering which can, in turn, lead to a vanishingly small direct-detection cross section.  It might also lead to isospin violation in the DM-nucleon scattering. Such non-SM values of light-quark Yukawa couplings may be probed in the high luminosity run of the LHC.
\end{abstract}

\maketitle

%%%%%%%%%%%%%%%%%%%%%%%%%%%%%%%%%%%%%%%%%%%%%%%%%%%%%%%%%%%%%%%%%%%%%%%%%%%%%%%%
\section{Introduction}
\noindent 
In the search for new physics (NP), a large class of current and future experiments are
dedicated to test the particle nature of dark matter (DM), especially by probing the ones known as the Weakly Interacting Massive
Particles (WIMPs) 
(see e.g.,~\cite{Jungman:1995df,Bertone:2004pz}). Among them, the  direct 
detection (DD) experiments rely upon the scattering of a DM particle off a detector nucleus, thereby looking for the recoiled nucleus. 
The list of current experiments in this direction includes Xenon1T~\cite{Aprile:2018dbl}, 
LUX~\cite{Akerib:2017kat}, PandaX-II~\cite{Cui:2017nnn}, SuperCDMS~\cite{PhysRevD.101.052008,
PhysRevD.99.062001} (also see~\cite{Undagoitia:2015gya}).
The  WIMP-nucleus scattering can be divided
into spin-independent (SI) and spin-dependent parts. Because of its coherent nature, the SI part is enhanced by a factor of $A^2$, where A is the mass number of the nucleus,
 making 
the SI scattering more relevant to the experiments
(see, e.g.,~\cite{Roszkowski:2017nbc,Schumann:2019eaa}). 

In the coming years, the Xenon-based detectors
{\sc LZ}~\cite{Akerib:2018lyp} or  XENONnT~\cite{Aprile:2020vtw}
are expected to give 
stringent bounds 
on the SI DD cross sections.
The projected sensitivity to probe SI WIMP-nucleon
scattering reaches to about
$1.4\times 10^{-12}$ pb for $M_{\rm DM}\sim40$-$50$ GeV.
For $M_{\rm DM}\sim1$ TeV, the upper bound on the
DM-nucleon scattering cross section goes to about $10^{-11}$ pb.
Obviously, such precise measurements would put strong  constraints
on the parameter spaces of several simple extensions of the 
Standard Model (SM), like the
$Z$-portal~\cite{Arcadi:2014lta,Hamaguchi:2015rxa,Escudero:2016gzx,Kearney:2016rng,Balazs:2017ple}, 
torsion-portal \cite{Barman:2019mlj},  Higgs-portal (H-portal)~\cite{Silveira:1985rk,McDonald:1993ex,Burgess:2000yq,Lebedev:2011iq,Djouadi:2011aa,Djouadi:2012zc,Cline:2013gha,
  Gross:2015cwa,Casas:2017jjg,Hoferichter:2017olk,Arcadi:2019lka},
$Z^\prime$-portal~\cite{Mambrini:2010dq,Dudas:2013sia,Alves:2013tqa,Lebedev:2014bba,Hooper:2014fda,
Alves:2015pea,Alves:2015mua,Allanach:2015gkd,Alves:2016cqf},
or the pseudoscalar-portals~\cite{Berlin:2015wwa,Baek:2017vzd,Bauer:2017fsw} etc. (see~\cite{Arcadi:2017kky} for a review).  
Among these, only 
the torsion-portal, $Z^\prime$-portal with only axial couplings and the pseudoscalar-portal models are somewhat 
preferred for fermionic DMs as the current null result in the SI DM-nucleon scattering can be accommodated. 
Otherwise, most popular TeV-scale DM models  
are already (or expected to be) under pressure from the existing (and future) SI DD limits, as the DM mass is pushed towards TeV values.

There have been some attempts to identify ways that would allow the WIMP scenarios to explain the non-observation of
any DM signal above the neutrino floor. 
The Higgs mediated models have drawn significant attention in this regard 
~\cite{Drees:1992am,Nath:1992ty,Baer:1997ai,Ellis:2003cw,Chattopadhyay:2008hk,
Chattopadhyay:2010vp,Das:2010kb,Chatterjee:2014bva,Feng:2010gw,Silveira:1985rk,McDonald:1993ex,Burgess:2000yq,Lebedev:2011iq,Djouadi:2011aa,Djouadi:2012zc,Cline:2013gha,
  Gross:2015cwa,Casas:2017jjg,Hoferichter:2017olk,Arcadi:2019lka}. 
They are relevant in many favoured beyond-the-SM (BSM) scenarios. 
The SM-like Higgs scalar can lead to large contributions at
the microscopic level through its coupling with the DM and quarks.
One may find parts of the
parameter spaces where the 
couplings of the DM to $Z$ or the Higgs boson may be highly suppressed or even zero identically. Similarly, in the
models with an extended Higgs sector, destructive interference between 
light and heavy CP-even Higgs exchanges  may lead to a cancellation in the SI scattering cross section. These are commonly known as the
``Blind spots'', since, in such cases,  the DD
experiments would not be able to probe the DM initiated recoils~\cite{He:2008qm,He:2011gc,Cheung:2012qy,Chang:2017gla,Huang:2014xua,Badziak:2015exr,
  Crivellin:2015bva,Han:2016qtc,Alanne:2017oqj,Altmannshofer:2019wjb,Alanne:2020xcb}.
Similarly, in a simple
H-portal dark-matter model, where a complex scalar is added to the SM, a softly broken symmetry might ensure that
the DM detection cross
section vanishes at the tree level for zero momentum transfer. Here, the
imaginary part of the complex scalar plays the role of the DM \cite{Gross:2017dan}. Ref. \cite{Balkin:2018tma} 
%contemplates a more general discussion on 
studies such a pseudo-Nambu-Goldstone WIMP DM in the context of composite Higgs models which provide with a compelling ultraviolet motivation. Isospin-violating DM (IVDM) \cite{Kurylov:2003ra,Giuliani:2005my,Chang:2010yk,Kang:2010mh,Feng:2011vu} is another interesting scenario where one can deal
with the same concern by allowing non-identical
couplings between the DM and nucleons. Since the limits
on the DM-nucleon scattering cross section assume isospin conservation, they cannot be directly used and hence, effectively become much relaxed~\cite{Feng:2011vu,Feng:2013vod,Yaguna:2016bga}.

In this paper, we pursue a {\it simple-minded} phenomenological analysis where a vanishing DM-nucleon scattering cross section can be  attained easily.
We look for a H-portal model that can account for the null result in the DD experiments.
However, instead of tuning the Higgs-DM interaction strength, we focus on the Higgs-nucleon interaction which is ultimately controlled by the Higgs-quark interactions, i.e., the Yukawa couplings of the SM quarks.
Here, we consider different signs and non-SM values of the quark Yukawa coupling(s) to make the net 
 SI  DM-nucleon/nucleus scattering cross section small. As an illustrative example, we
consider the 
possibility of a singlet scalar DM.
Earlier, it has been shown that the 
Yukawa interactions of the SM quarks leave room for a partial cancellation in their contribution to the Higgs-nucleon coupling within
the two-Higgs-doublet extension of the SM along with a real singlet scalar DM~\cite{He:2008qm,He:2011gc}. 
Implications of non-SM-like Higgs Yukawa
couplings to the light
quarks on the  H-portal DM phenomenology have also been studied in Ref.~\cite{Bishara:2015cha} where the authors showed that the
DD scattering rate can increase by up to four orders
of magnitude. Here, in particular, we will show that in NP models, it is possible to have partial or even complete cancellation of the Higgs-nucleon couplings in the presence of higher dimensional operators at the TeV scale. Another potential benefit can be found, since the DM-nucleon effective coupling may become isospin-violating,  
in general. This conclusion can be generalized to 
two Higgs doublet models including the Supersymmetric (SUSY) ones.

The main difficulty in modifying the $hq\bar q$ interactions lies within the quark masses --- in the SM, the
Yukawa couplings of the fermions are proportional to the respective masses, i.e., 
$y_q \propto m_q/v$, where $v$ is the vacuum expectation value (VEV) of the SM Higgs. Any
deviation coming from the higher-order terms, particularly for the light fermions,
are suppressed and unobservable
in the SM. However,
in the presence of new physics 
at the TeV scale,
such alignments between the quark masses and Yukawa matrices
can be relaxed (see~\cite{Buchmuller:1985jz,Grzadkowski:2010es} for discussions within the context of effective operators and~\cite{Bhaskar:2020kdr}, for the radiative generation of such a higher dimensional operator). 
Hence, the quark Yukawa couplings to the
physical Higgs boson $h$, especially for the light ones ($u,d,c,s$),  may take any value 
permissible by the measurements at the LHC. 
For the third-generation fermions and the vector bosons, these couplings
are rather tightly constrained and have already been measured within
10\%-20\% of their SM 
predictions at the LHC~\cite{Aad:2019mbh}. But, the same for the first two generations
of light quarks are not yet 
measured directly --- significant deviations from their SM values are still allowed~\cite{deBlas:2019rxi,Falkowski:2020znk} and can offer
interesting phenomenology~\cite{Hedri:2013wea,Egana-Ugrinovic:2018znw,Egana-Ugrinovic:2019dqu,Bar-Shalom:2018rjs,Bhaskar:2020kdr}.
Interestingly, for the light quarks, negative Yukawa couplings are also allowed~\cite{Aad:2019mbh,deBlas:2019rxi} and have drawn 
some interests in recent times~\cite{Bishara:2016jga,Bonner:2016sdg}.
This, in the present
context, is highly welcome as it allows 
the SI DM-nucleon scattering cross section to become vanishingly
small through a cancellation among the different quark contributions to the Higgs-nucleon interaction.
Note that our results are valid for
any TeV scale DM model with the Higgs scalar as the dominant source for SI scatterings, since the small SI DM-nucleon scattering cross section does not depend on
how the DM interacts with the Higgs scalar.

This paper is organized as follows. We first review the prerequisites for calculating the SI DM DD
cross section in Sec.~\ref{sec:SIhiggs}. Then in Sec.~\ref{sec:SMHiggs}, we discuss the general framework 
where the quark Yukawa couplings to the physical Higgs field can
be attuned. We illustrate how a specific type of higher dimensional operator could help. Then we discuss a
TeV scale extension of the SM with some vector-like (VL) quarks where the desired dimension-6
operator is generated effectively at the tree level.
A real scalar is assumed to be the DM in this model. We also discuss the existing LHC constraints and the possibility of 
probing the required non-SM light-quark Yukawa couplings at the colliders. In Sec.~\ref{sec:SI}, we present our numerical results. There
we consider the Yukawa couplings as free parameters and show how a
negative coupling (in this case, of the second-generation quarks) can be helpful
in explaining the null results in the DD experiments. Finally,  we conclude in Sec.~\ref{sec:conclu}.

%%%%%%%%%%%%%%%%%%%%%%%%%%%%%%%%%%%%%%%%%%%%%%%%%%%%%%%%%%%%%%%%%%%%%%%%%%%%%%%%
\section{Spin Independent Direct Detection Cross Section: Higgs Exchanges}
\label{sec:SIhiggs}
\noindent
%Neglecting the nuclear structure effects,
For a real scalar dark matter $\phi$, the SI DM-nucleon elastic scattering cross section at zero momentum transfer 
can be written as, 
\begin{align}
\sigma^{\phi-N}_{\rm SI}=\frac{m_r^2 f_N^2}{4 \pi M^2_{\rm \phi}},
\label{eq:sigma_SI}
\end{align}
%\begin{align}
%\sigma_{\rm SI}=\frac{m_r^2}{4 \pi M^2_{\rm \phi}}\left[Zf_p+(A-Z)f_n\right]^2,
%\label{eq:sigma_SI}
%\end{align}
%where, A and Z are the mass number and the atomic number of the nucleus,
%respectively, 
where, $M_\phi$ is mass of the DM and $m_r$
is the DM-nucleon reduced mass. 
At the microscopic level,
the effective DM-quark scattering can be read from the following interaction term:
\begin{align}
\mathcal{L}_q^{\rm SI}=f_q(\phi\phi)(\bar{q}q),
\label{eq:opquark}
\end{align}
where $q$ denotes a quark and $f_q$, the corresponding scattering coefficient.
In a simple model of H-portal DM, we can set
$f_q = \lambda_{\phi}\,y_q/m_H^2$ where
$y_q$ is the Yukawa coupling of the quark $q$ and
$\lambda_{\phi}$ is the effective DM coupling to the SM Higgs boson. Similarly, the DM-nucleon interaction can be expressed as,
\begin{align}
\mathcal{L}_N^{\rm SI}=f_N(\phi\phi)(\bar{N}N),
\label{eq:opnucleon}
\end{align}
where $f_N = \lambda_{\phi}\,\lambda_N/m_H^2$ and
$\lambda_N = m_N \sum_{q} (y_q f^{(N)}_{q}/m_q)$ with $m_N$
being the nucleon mass.
The effective 
interaction of the DM particle with a nucleon (proton or neutron) can be
obtained from the expectation
value of the operator in Eq.~\eqref{eq:opquark} with respect to the initial and final nucleon states
($N \equiv p$ or $n$) \cite{SHIFMAN1978443}. Here, one may use the fact that nucleon mass is determined from the trace of the energy-momentum tensor. 
Generically, for $q=u,d,s$ the factor
$f^{(N)}_{q}$  can be expressed as
\begin{align}
\langle N|m_q\bar{q}q|N\rangle=m_Nf^{(N)}_{q}.
\end{align} 
For the heavier quarks, $f^{(N)}_{q}$ can be evaluated through one-loop contributions due to scattering off gluons~\cite{SHIFMAN1978443,Drees:1993bu}:
\begin{align}
 f^{(N)}_{c,b,t} = \frac2{27}f^{(N)}_{G} = \frac2{27}\left(1-\sum_{q=u,d,s}f^{(N)}_{q}\right).
\end{align}
In fact, to the leading order,
the effective $H G^{\mu\nu}G_{\mu\nu}$ vertex at small momentum transfer can also be
used for the above  computation~\cite{Djouadi:2000ck}. The dominant QCD corrections should
also be taken into account in that case.  
Altogether, one can cast the effective DM-nucleon scattering coefficient as~\cite{Belanger:2008sj,Arcadi:2019lka},
\begin{align}
  \frac{\lambda_N}{m_N}=\sum_{q=u,d,s}f^{(N)}_{q}\frac{y_q}{m_q}+\frac{2}{27}f^{(N)}_{G}\sum_{q=c,b,t}\frac{C_q y_q}{m_q}.
\label{eq:FN}
\end{align}
The parameters, $f^{(N)}_{q}$ ($q \in u,d,s$)               
can be determined from lattice QCD calculations~\cite{Thomas:2012tg}. For the heavier quarks, the leading order QCD correction becomes $C_q=1+11\alpha_s(m_q)/{4\pi}$. 
We use the following values of $f^{(N)}_{q}$~\cite{Thomas:2012tg,Belanger:2013oya},
\begin{align}
&f^p_{u} =0.0153,\quad f^p_{d} =0.0191,\quad f^p_{s} =0.0447, \nonumber\\
&f^n_{u} =0.0110,\quad f^n_{d} =0.0273,\quad f^n_{s} =0.0447
\label{eq:ff}
\end{align}
which lead to $f^{(N)}_G \sim 0.92$ (ignoring the
differences between nucleons $N=p,~n$).\footnote{{One gets slightly different values from chiral perturbation theory~\cite{Alarcon:2011zs,Crivellin:2013ipa,Hoferichter:2015dsa,Arcadi:2019lka}.}}
It should be noted that the above numerical values are subject to some uncertainties as
they are evaluated using the hadronic data.

As indicated in the Introduction, we allow
the light-quark Yukawa couplings to deviate from their
respective SM values, and more importantly, allow them to attain negative values which are not in violation with any experimental
observation so far. We use this freedom to delineate the tentative range for 
$y_s$ or $y_c$ in terms of the other SM-like Yukawa couplings so that
$\lambda_N = 0$ can be achieved. Since $C_q$ is very close to $1$, here, we may simply assume $C_q = 1$  to get a qualitative picture. Thus, by substituting the values for
$f^{(N)}_q$ in Eq.~\eqref{eq:FN}, one would get a typical
regime for the Yukawa couplings where $\lambda_N$ would be vanishingly
small, i.e., 
\begin{align}
  y_s = -\frac{m_s}{f^{(N)}_s}\left(f^{(N)}_u\frac{y_u}{m_u}+f^{(N)}_d\frac{y_d}{m_d}\right);
   \quad
  y_c = -m_c\left(\frac{y_b}{m_b}+\frac{y_t}{m_t}\right).
  \label{eq:nulDD_anal}
\end{align}
In this illustrative example, we have allowed only $y_s$ and $y_c$ to take non-SM values for simplicity and further assumed that 
$y_s$ cancels the light quark contribution and $y_c$ cancel the heavier quark contribution separately. Obviously, this assumption is only a choice that we make for the illustration and not a requirement. One may tune any one or more light-quark Yukawa couplings to attain $\lambda_N=0$.
The
above equation simplifies to
\begin{align}
 y_s = -0.770\, y^{\rm SM}_s ~; \quad
  y_c = -2\, y^{\rm SM}_c~,
  \label{eq:analyukawa}
\end{align}
for $N=p$. Notably, for this set of parameters,
$\lambda_p \neq \lambda_n$, implying that the DM-neutron scattering cross section will not vanish identically, and hence, a degree of isospin violation would be observed. 
Furthermore, these changes are potentially insignificant to modify the effective $ggh$ vertex in the SM.

We will further discuss about this possibility in Sec.~\ref{sec:SI}. A similar cancellation condition can be achieved for the neutron as well by using the respective form factors where one finds $ y_s = -0.857\, y^{\rm SM}_s $. 
If one only allows either $y_c$ or $y_s$
to take values such that the DM-nucleon effective coupling in
Eq.~\eqref{eq:FN} vanishes, then a relatively larger negative values would be required, which may need some fine-tuning. Similarly, $y_u$ and/or $y_d$ can also be considered
to be negative to achieve the same result.

%%%%%%%%%%%%%%%%%%%%%%%%%%%%%%%%%%%%%%%%%%%%%%%%%%%%%%%%%%%%%%%%%%%%%%%%%%%%%%%%
\section{Non-SM Light-quark Yukawa Couplings: Examples and Experimental
 Tests}
\label{sec:SMHiggs}

\noindent
In this section, we illustrate how the non-SM-like Yukawa couplings can be generated through
dimension-6 operators at the tree level in an effective theory framework. The current LHC measurements prevent large variations in $y_{t,\,b}$ \cite{PhysRevD.101.012002} but leave space for deviations in the Yukawa couplings 
of the first two generations of quarks (for large changes to the top Yukawa couplings, see \cite{Hedri:2013wea}).
A discussion on the collider tests of
non-SM values of $y_q$ follows in the next section.

%%%%%%%%%%%%%%%%%%%%%%%%%%%%%%%%%%%%%%%%%%%%%%%%%%%%%%%%%%%%%%%%%%%%%%%%%%%%%%%%
\subsection{Non-SM-like Light-quark Yukawa Couplings and Higher Dimensional Operators}

\noindent 
In this example, we include a particular type of effective dimension-6 operators at some NP scale $\Lambda$ in the quark Yukawa interaction Lagrangian,
\begin{align}
  \mathcal{L} \supset -Y_u\bar{q}_L\tilde{H}u_R-Y_d\bar{q}_LHd_R+ \Delta\mathcal{L}_{eff} +H.c.,
  \label{eq:lageff}
\end{align}
where
\begin{align}
\Delta\mathcal{L}_{eff}=\frac{H^{\dagger}H}{\Lambda^2}\left(Y_H^{u}\bar{q}_L\tilde{H}u_R+Y_H^{d}\bar{q}_LHd_R\right).
\label{eq:DL1}
\end{align}
Here, $Y_H^{u,\,d}$ are the Wilson
coefficients determined by the details of the NP model. In general, $Y_q$ and $Y_H^{q}$ are assumed to be $3 \times 3$ matrices
in the generation space. The SM Higgs doublet is denoted by
$H$ ($\tilde{H}\equiv i\tau_2H^*$), $q_L$ is the left-handed
$SU(2)$ quark doublet and $u_R$, $d_R$ are the right-handed up- and down-type quarks, respectively.
After the electroweak symmetry breaking (EWSB),   using $H=\left(\begin{array}{c}{0}\\{h+v}\end{array}\right)$
with $v \simeq 174$ GeV, one obtains the
quark mass matrix ${M}_q$ and the corresponding Yukawa coupling matrix
in the mass basis
by considering unitary rotations of both left- and right-handed quark
fields. For simplicity, we ignore flavour mixings and assume diagonal 
NP Yukawa couplings, i.e.,
$Y_H^{q} = \mathbb{I}_{3 \times 3}$ in the same basis. Though this is not
true in general, it suffices for the present purpose.
Now, we see that for a quark ($q$), its physical mass ($m_q$) and 
Yukawa coupling ($y_q$) to the physical Higgs boson ($h$) become non-aligned, i.e., 
\begin{align}
m_q&=v\left(Y_q-\epsilon Y_H^{q}\right), \label{eq:yq}\\
y_q&=\left(Y_q-3\epsilon Y_H^{q}\right)
=\frac{m_q}{v}-2\epsilon Y_H^{q}
\label{eq:mYq}
\end{align}
where, $\epsilon\equiv\left(v/\Lambda\right)^2$. 
Assuming
$\Lambda \sim$ TeV and $Y_H^{q} \simeq \mathcal{O}(1)$, a few comments are in order. 
\begin{itemize}
\item
  It is clear that, when a higher-order operator [like the one shown in Eq.~\eqref{eq:DL1}] is added to the SM Lagrangian, 
  the fermion mass and Yukawa coupling become two independent 
  quantities in the physical basis. In other words, it gives
  us the freedom to modify the quark-Higgs Yukawa couplings without 
  perturbing the quark masses.
  \item
The sign
of the Yukawa couplings $y_q$ depends
on  the  sign  of  the  Wilson  coefficients $Y_H^{q}$.  In  particular, for the first
two generations of quarks $(u,d,s,c)$, one may find that 
$m_q/v \ll \epsilon Y_H^{q}$, thus the respective Yukawa couplings may
naturally become negative.
\item
To achieve the correct size and sign of $m_q$ with a negative $y_q$ [see e.g., Eq.~\eqref{eq:nulDD_anal}], one may use Eqs.~\eqref{eq:yq} and~\eqref{eq:mYq} to obtain,
\begin{align}
Y^q_H\left(\frac{v}{\Lambda}\right)^2 >\frac{m_q}{2v}.
\label{eq:lambda_bound}
\end{align}
This sets an upper bound on the NP scale $\Lambda$ --- for
a perturbative choice of $Y^q_H \sim \mathcal O(1)$,
the NP scale $\Lambda$ should not be much larger than a few TeV. For example, Eq.~\eqref{eq:lambda_bound} implies that $\Lambda$ should be lower than about $2.9$ TeV if we take the charm quark (i.e., set $m_q=m^{\rm SM}_c$ and $Y^q_H=1$). On the
contrary, $y_q > 0$  can only set a lower bound on $\Lambda$. Thus the choice of negative values for 
the Yukawa couplings
is more natural and predictive compared to the positive values. 

\item
  Usually, if one considers with full generality,  $Y_q$ and $Y_H^{q}$ cannot be
  diagonalized simultaneously in the mass basis.  As a result, Higgs mediated flavour
  changing neutral couplings among the SM quarks, which are
  otherwise extremely constrained by experiments, may appear. However, such flavour
  changing couplings
may be suppressed if a definite flavour symmetry or some flavour selection rules are applied \cite{Bar-Shalom:2018rjs}. Since it hardly has any impact in the present case, we will not discuss it anymore.
\end{itemize}

%%%%%%%%%%%%%%%%%%%%%%%%%%%%%%%%%%%%%%%%%%%%%%%%%%%%%%%%%%%%%%%%%%%%%%%%%%%%%%%%
\subsection{Singlet Scalar DM and Negative Light-quark Yukawa Couplings}\label{sec:vlq}

\noindent
Here, we consider a specific realization where the aforesaid
dimension-6 operators are generated through some underlying NP that includes
new heavy VL particles. VL fermions have a rich phenomenology and can be found 
in various popular NP scenarios, e.g., in some 
SUSY extensions~\cite{Kang:2007ib,Babu:2008ge,Graham:2009gy,Martin:2010dc}, composite Higgs models~\cite{Contino:2006qr,Anastasiou:2009rv,Vignaroli:2012sf,DeSimone:2012fs}, warped extra-dimension models~\cite{Agashe:2003zs,Agashe:2004bm,Agashe:2004cp,Contino:2008hi,Gopalakrishna:2011ef,Gopalakrishna:2013hua},
little Higgs models~\cite{Han:2003wu,Carena:2006jx,Matsumoto:2008fq,Berger:2012ec}, etc.

In general, the effective Lagrangian can be considered to have three parts:
\begin{align}
\mathcal{L}_{eff} = \mathcal{L}_{\rm SM} + \mathcal{L}_{\rm NP} + \mathcal{L}_{\rm DM}
  \end{align}
where $\mathcal{L}_{\rm SM}$ is the usual SM part, $\mathcal{L}_{\rm NP}$ includes the
new physics terms (except the DM part) responsible mainly for the nonalignment of 
the Yukawa couplings of
the physical Higgs boson to SM quarks once the heavy fields are integrated out, 
and $\mathcal{L}_{\rm DM}$ refers to the terms involving DM interactions. Usually, these three parts may couple 
with each other in specific scenarios. For example, here
we will see the NP-SM
couplings producing the necessary deviations in Yukawa couplings at the tree level.

As mentioned in Sec.~\ref{sec:SIhiggs}, we consider a real singlet scalar $\phi$ as the DM particle while the
SM Higgs acts as the portal.
An additional  $Z_2$ symmetry ensures the stability of $\phi$. Apart from the usual kinetic term in $\mathcal{L}_{\rm SM}$, the potential term 
for the model can be written as,
\begin{align}
V= \frac{1}{2}\mu_{\phi}^2\phi^2+\lambda_{H\phi}(H^{\dagger}H)
\phi^2.
\label{eq:DM_pot}
\end{align}  
In principle, there should be a self-interacting $\phi^4$ term also. However, we can safely
ignore it, since it will have no effect  in determining either the relic
density or the DD cross section. 
After EWSB, $\phi$ doesn't get any VEV. Hence, the $\phi$-mass term can
be defined as,
\begin{align}
M_{\phi}=\sqrt{\mu_{\phi}^2+2\lambda_{H\phi}v^2}.
\label{eq:DM_mass}
\end{align}
The interaction terms connecting the dark sector with the
SM are $(2\lambda_{H\phi}v)h\phi^2$ and $\lambda_{H\phi}h^2\phi^2$. The phenomenology of this model has been widely studied.
But as discussed earlier, it is quite
difficult to satisfy the current direct detection bounds with this simple
extension and this motivates us to propose $\mathcal{L}_{\rm NP}$. 

In ${\cal L}_{\rm NP}$, we introduce the VL fermions and argue that the non-SM-like Yukawa couplings 
for the light quarks 
can be generated at the tree level once the heavy VL fermion degrees are integrated out.  
Based on our previous discussion it is clear that we want non-SM-like Yukawa couplings,
specifically for the $2^{\rm nd}$ generation quarks to the SM Higgs scalar.
Thus, we consider a
SM-like set up to specify the underlying NP theory 
\cite{Bar-Shalom:2018rjs,Hedri:2013wea}
that includes only one generation of VL quarks: an $SU(2)$ VL quark doublet $Q = (C,S) (3,2,1/6)$ and the corresponding
up-type and down-type $SU(2)$ singlets $C (3,1,2/3)$ and
$S (3,1,-1/3)$, carrying
the same quantum numbers as the SM quark doublets and singlets.
Further, we assume the VL quarks are in their mass basis, 
with $M_{Q,\,C,\,S} \gtrsim 2$ TeV, putting them well above the current LHC bounds. In general, we may
write down the NP Lagrangian for the interactions among 
the VL quarks and the SM states, as follows.   
\begin{align}
- {\cal L}_{\rm NP} =&  \quad\left(\lambda_{QC}\, \bar Q_L \tilde H C_R +
\lambda_{QS}\, \bar Q_L H S_R \right) 
+
\left(\lambda_{qC}\, \bar q_L \tilde H C_{R} +
\lambda_{qS}\, \bar q_L H S_{R}\right) \nonumber \\
&+ \left(\lambda_{Qc}\, \bar Q_L \tilde H c_R +
\lambda_{Qs}\, \bar Q_L H s_R\right) + H.c.\label{NPmodel1}
\end{align}
Here $q_L$ refers to the second-generation SM quarks. 
This Lagrangian can lead to the desired dimension-6 operators in Eq.~\eqref{eq:DL1} after integrating out heavy the VL quarks. The dimension-6 couplings
$Y^{c}_{H}, Y^{s}_{H}$, as defined in Eq.~\eqref{eq:DL1}, 
can be obtained from the diagrams shown in Fig.~\ref{fig:model1figs} and are given below:\footnote{In general, the NP couplings 
are all $3 \times 3$
matrices in the flavour space, though for the present purpose, it suffices to
assume 
$\lambda_{NP} = \mathbb{I}_{3 \times 3}$.
The
SM quark fields are also assumed to be in their physical mass basis. Usually,
with the VL quarks the CKM matrix is
extended and the SM $3 \times 3$ CKM block is, in principle, no longer unitary,
though the deviation is marginal \cite{Bar-Shalom:2018rjs}.
This results into
$3 \times 3$ diagonal matrices for the 
Wilson coefficients $Y^{u_i}_{H}, Y^{d_i}_{H}$ in Eq.~\eqref{eq:DL1}.}
\begin{eqnarray}
Y^{c}_{H} = \lambda_{qC}\lambda_{QC}^* \lambda_{Qc} ~&,&~
\Lambda = \sqrt{M_C M_Q} ~, \label{Yuh} \\
Y^{s}_{H} = ~\lambda_{qS} \lambda_{QS}^* \lambda_{Qs} ~&,&~
\Lambda = \sqrt{M_S M_Q} \label{Ydh} ~.
\end{eqnarray}
Thus, if all the VL quarks have masses
$M_C \sim M_S \sim M_Q \sim 2$ TeV and the
 new physics couplings $\lambda_{NP} \sim {\cal O}(1)$, 
the Yukawa couplings
of the second-generation quarks can be considered for 
modification, as shown in Eq.~\eqref{eq:mYq}.
\begin{figure}[!t]
\subfloat[(a)\quad\quad]{\includegraphics[scale=0.4]{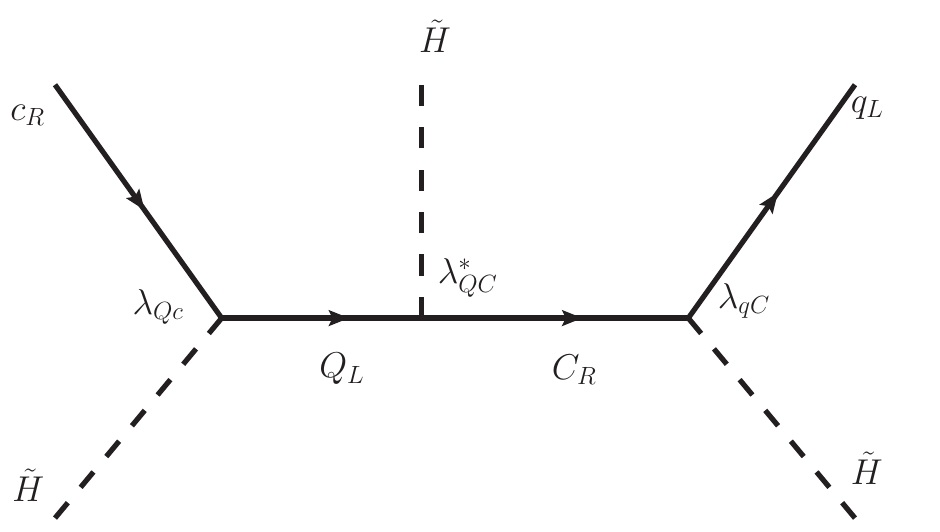}}\quad
\subfloat[(b)\quad\quad]{\includegraphics[scale=0.4]{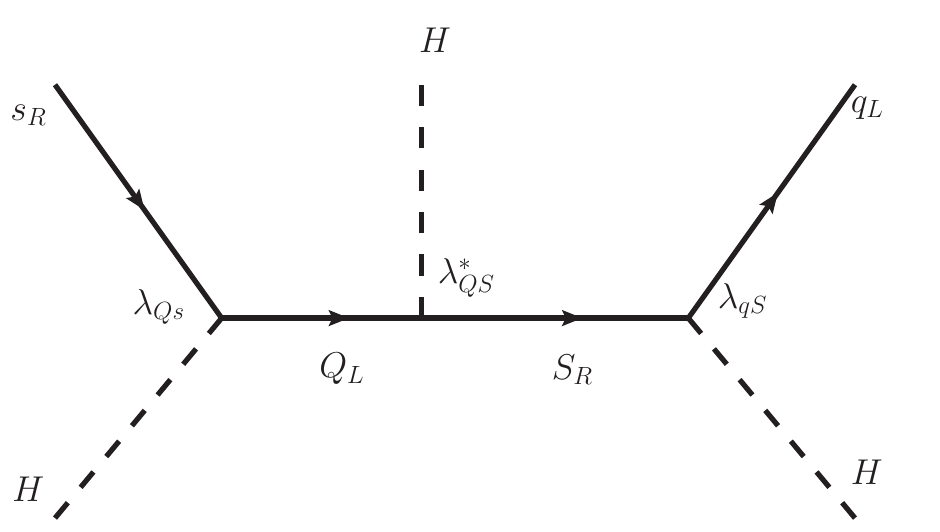}}
\caption{Dimension-6 operators relevant for (a) $\tilde{H}^\dagger \tilde{H} c_R\rightarrow q_L\tilde{H}$ and (b)
  $H^\dagger H s_R\rightarrow q_L H$ process initiated at the tree level after
  integrating out heavy VL quarks.}
\label{fig:model1figs}
\end{figure}
%
%%%%%%%%%%%%%%%%%%%%%%%%%%%%%%%%%%%%%%%%%%%%%%%%%%%%%%%%%%%%%%%%%%%%%%%%%%%%%%%%
\subsection{Tests of the Non-standard Light-quark Yukawa Couplings at the LHC and Beyond}

\noindent 
The SM Higgs couplings to
the massive vector bosons, the third generation quarks and the $\tau$-lepton and its effective couplings to two gluons or two photons are known with some accuracy (roughly $10\%-20\%$)~\cite{Aad:2019mbh}. Recently the ATLAS and CMS collaborations announced the first observation of a Higgs decaying to two $\mu$'s~\cite{Aad:2020xfq,Sirunyan:2020two}. However, how the Higgs couples to the other light fermions has never been probed experimentally --- the LHC is yet to observe a direct Higgs decay to a pair of electrons or any of the first two generation of quarks. 

Because of the Higgs mechanism, the Yukawa coupling of a fermion is proportional to its mass in the SM. This makes  the Yukawa couplings of the light fermions harder to probe. However, with the high luminosity run of the LHC (HL-LHC), there could be some chance of probing them. In recent times, several possibilities have been explored for testing these couplings at the LHC~\cite{Bodwin:2013gca,Delaunay:2013pja,Kagan:2014ila,Perez:2015aoa,Perez:2015lra,Koenig:2015pha,Brivio:2015fxa,Bishara:2016jga,Soreq:2016rae,Bonner:2016sdg,Yu:2016rvv,Cohen:2017rsk,Han:2018juw,Mao:2019hgg,Coyle:2019hvs,Alasfar:2019pmn,Aguilar-Saavedra:2020rgo}, especially for $c$ and also for $s$-quarks, in parallel to the ongoing experimental efforts~\cite{Aad:2015sda,Aaboud:2016rug,LHCb:2016yxg,Aaboud:2017xnb,Aaboud:2018fhh,Sirunyan:2020mds}. 
Some of these proposals consider looking at rare Higgs decays to light flavoured mesons (like $J/\psi$ or $\phi$, etc.)~\cite{Bodwin:2013gca,Kagan:2014ila,Koenig:2015pha, Mao:2019hgg}. Even though these can offer clean signals, such strategies suffer from low signal rates due to the small decay rates involved. There are channels with relatively larger signal cross sections [like $pp \to W/Zh \to W/Z (cc)$~\cite{Delaunay:2013pja,Perez:2015lra,Perez:2015aoa}, $pp \to hc$ \cite{Brivio:2015fxa}, $pp\to hh \to (cc)(\gamma\gamma)$~\cite{Alasfar:2019pmn} etc.] that require  light-quark jet tagging. Ref.~\cite{Han:2018juw} uses a refined triggering strategy and some machine learning techniques to probe $pp\to h\to c\bar c\gamma$ and estimates the HL-LHC reach as $\left|y_c/y_c^{\rm SM}\right|<8$.  The process $pp\to h\gamma$ is used in Ref.~\cite{Aguilar-Saavedra:2020rgo}. Ref.~\cite{Yu:2016rvv}  points out the possibility of using the charge asymmetry in  $pp\to W^\pm h$ to constrain the 
light-quark Yukawa couplings. Ref.~\cite{Coyle:2019hvs} employs a combination of the above ideas. Ref.~\cite{Cohen:2017rsk} looks at the  Higgs $p_{\rm T}$ distribution in $pp\to h+j(j_b)\to\gamma\gamma +j(j_b)$. In the gluon-gluon fusion process, the Higgs is mostly produced centrally, i.e., about $y_h\approx 0$.  For a non-negligible $y_u$, the Higgs would also be produced via $u\bar u$ fusion. However, since only $u$ is a valance quark of proton, not $\bar u$, production from the $u\bar u$ fusion would peak in forward region, i.e., around higher $|y_h|$. Hence, Ref.~\cite{Soreq:2016rae} considers the idea of using both $p_{\rm T}$ and rapidity distributions of the Higgs to obtain bounds on the light Yukawa couplings. Ref.~\cite{Falkowski:2020znk} considers triple heavy vector boson production as a probe of the first-generation light-quark Yukawa couplings.
These proposals are somewhat competitive in nature as the projected limits obtained for either the $300$ fb$^{-1}$ LHC or HL-LHC are similar. Here, as an estimation, we quote the projected reach in the
$y_q$ values ($q=u,\,d,\,s,\,c$) at the LHC with  $3000$ fb$^{-1}$ of integrated luminosity from Ref~\cite{deBlas:2019rxi}:
\begin{align}
  |y_u|<560~y^{\rm SM}_u, 
  \quad|y_d|<260~y^{\rm SM}_d, 
  \quad|y_s|<13~y^{\rm SM}_s, 
  \quad|y_c|<1.2~y^{\rm SM}_c.
\end{align}

 Notice that the limits are put on the absolute values of the Yukawa couplings as the processes involved are insensitive to the sign of the light-quark Yukawa couplings. There are, however, some processes that have some sensitivity on the signs. In the process where a Higgs is produced with a jet ($pp\to hj$), the shape of the $p_{\rm T}$ distribution of the Higgs depends on the production mode. Ref.~\cite{Bishara:2016jga} utilizes this to estimate that the HL-LHC can restrict $y_c/y_c^{\rm SM}$ to be within $[-0.6,3]$. Here, the interference between $c$- and $t$- quark loops gives rise to a term linear in $y_c$ in the cross section of the $gg\to hj$ subprocess, making it somewhat sensitive to the sign of $y_c$. 
Ref.~\cite{Bonner:2016sdg} also considers the $p_{\rm T}$ distribution of Higgs. They consider the process $pp\to h\to 4\ell$ at the next-to-leading order (NLO) where the quark fusion and gluon fusion subprocesses can interfere. For the HL-LHC, they find $-1550<y_u/y_u^{\rm SM}<700$ and $-800<y_d/y_d^{\rm SM}<300$. 

There are other ways to probe the light-quark Yukawa couplings than at the LHC or HL-LHC. For example, Ref.~\cite{Goertz:2014qia} indicates that FCNC transitions in the Kaon sector could restrict the down-quark Yukawa coupling to be within the rage $0.4 < |y_d/y_d^{\rm SM}| < 1.7$. It is possible to probe the light-quark Yukawa couplings by measuring isotope shifts, that are affected by Higgs exchange, in atomic clock transitions~\cite{Delaunay:2016brc}. Ref.~\cite{Bishara:2015cha} 
points out that a discovery of H-portal dark matter could let us put bounds on these couplings. The future colliders, especially the leptonic ones, could also offer us a better handle in measuring the Higgs couplings in general~\cite{deBlas:2019rxi}. This is mainly because the lepton colliders are clean and allow us to reconstruct the processes far more accurately than their hadron counterparts. For example, Ref.~\cite{Gao:2016jcm} considers probing  via hadronic event shapes at lepton colliders and shows that light-quark Yukawa couplings greater than $|0.09\times y_b^{\rm SM}|$ might be excluded in an $e^+e^-$ collider of  centre-of-mass energy $250$ GeV with an integrated luminosity of $5$ ab$^{-1}$.

In the subsequent analysis, we compute $\sigma_{\rm SI}$ with non-SM values of $y_c$ and $y_s$,
especially with negative values of $y_c$ and $y_s$, 
so that the SI scattering cross section may become vanishingly small. 
Hence, the future generation experiments like XENONnT or LZ could only assert our proposal through
their blindness to find any signal in the $\sigma_{\rm SI}$-$M_{\rm DM}$ plane. But,
on the other hand, parts of the parameter space with non-SM values for
$y_c$ (this includes the negative values as well) 
can be tested at the HL-LHC as claimed in Refs.~\cite{Bishara:2016jga,deBlas:2019rxi}. This 
complementarity between the DM and LHC searches might help in testing our
proposal.

%%%%%%%%%%%%%%%%%%%%%%%%%%%%%%%%%%%%%%%%%%%%%%%%%%%%%%%%%%%%%%%%%%%%%%%%%%%%%%%%
\section{Relic Density and Direct Detection of DM}
\label{sec:SI}

\begin{figure}[!t]
\begin{center}
  \includegraphics[scale=0.5]{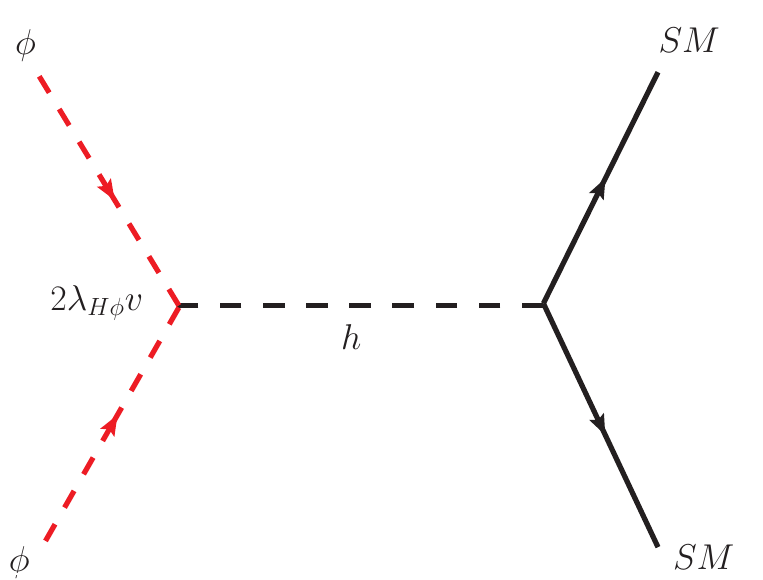}
\caption{Dark matter annihilation into
  SM particles through Higgs mediated $s$-channel process. In $t$-channel, this diagram contributes to
  $\sigma_{\rm SI}$.}
\label{annihilation}
\end{center}
\end{figure}

\noindent For numerical analysis, we use the code 
micrOMEGAs~\cite{Belanger:2006is,Belanger:2008sj} to evaluate the relic density and
DD cross section. The dominant QCD corrections in the SI DM-nucleon
scattering are already included in the code.
In our model, the main two
free parameters  are $M_\phi$ and $\lambda_{H\phi}$. Additionally,
we consider $y_c$ and $y_s$ also as free parameters.
In our computation, 
we set $\lambda_{H\phi}=0.02$.
 The valid parameter space should
 comply with the observed relic abundance data ~\cite{Jarosik_2011,Aghanim:2018eyx},
 \begin{equation}
   \Omega_{\rm DM}h^2=0.1198\pm 0.0012.
   \label{eq:relic}
 \end{equation}
For the singlet scalar DM $\phi$, one can easily solve the Boltzmann equation to get the corresponding relic abundance. The Boltzmann equation is given by,
\begin{align}
\frac{dn}{dt} + 3 \mathcal H n = - \langle \sigma_{eff}\, v \rangle (n^2 - n_{eq}^2) 
\label{eq:Boltz}
\end{align}
where $\mathcal H$ is the Hubble constant and $\langle \sigma_{eff}\, v \rangle$ is the thermal averaged cross section of the DM annihilation to the SM particles. In this scenario, only the $s$-channel process shown in Fig.~\ref{annihilation} keeps the DM in thermal equilibrium.
\begin{figure}[!t]
\begin{center}
\includegraphics[scale=0.65]{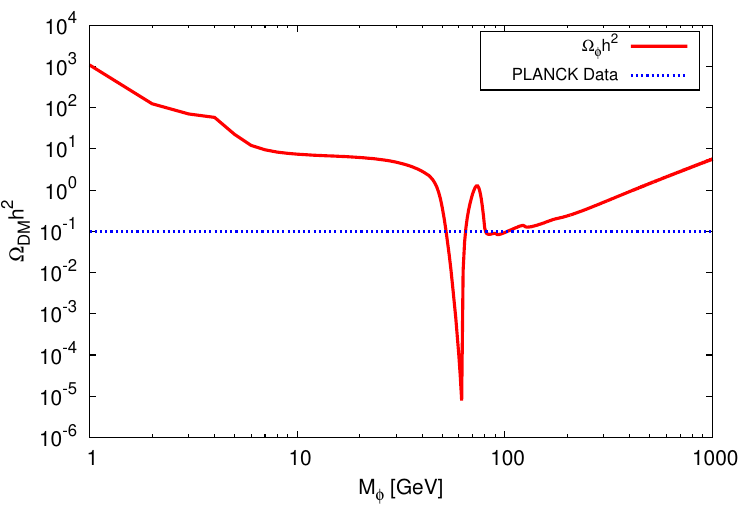}
\caption{Variation of relic abundance as a function of $M_{\phi}$.}
\label{fig:Rplot}
\end{center}
\end{figure} 
\begin{figure}[!t]\begin{center}
\subfloat[\quad\quad\quad(a)]{\includegraphics[scale=0.6]{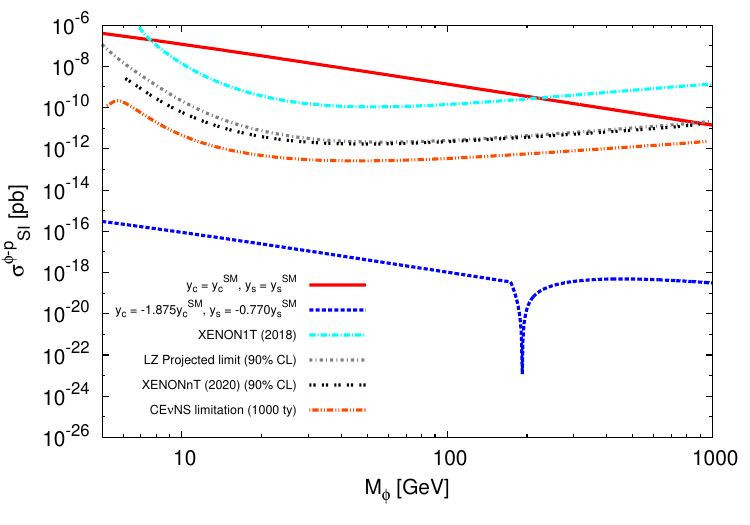}\label{fig:CrossA}}\quad
\subfloat[\quad\quad\quad(b)]{\includegraphics[scale=0.6]{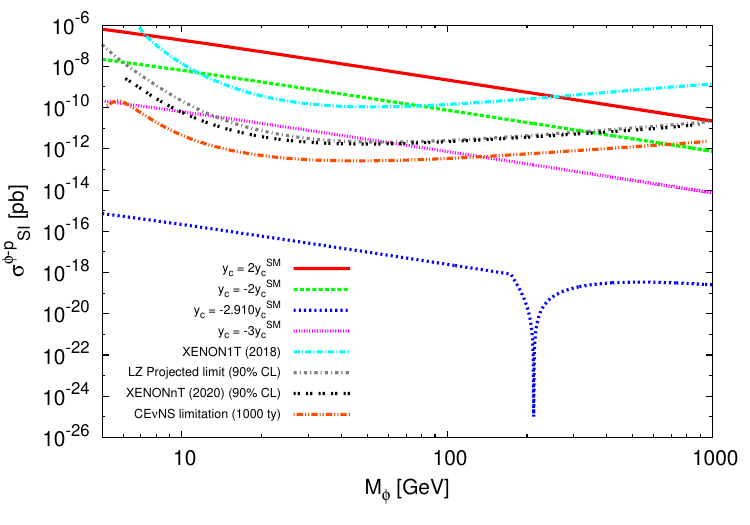}\label{fig:CrossB}}\\
\subfloat[\quad\quad\quad(c)]{\includegraphics[scale=0.6]{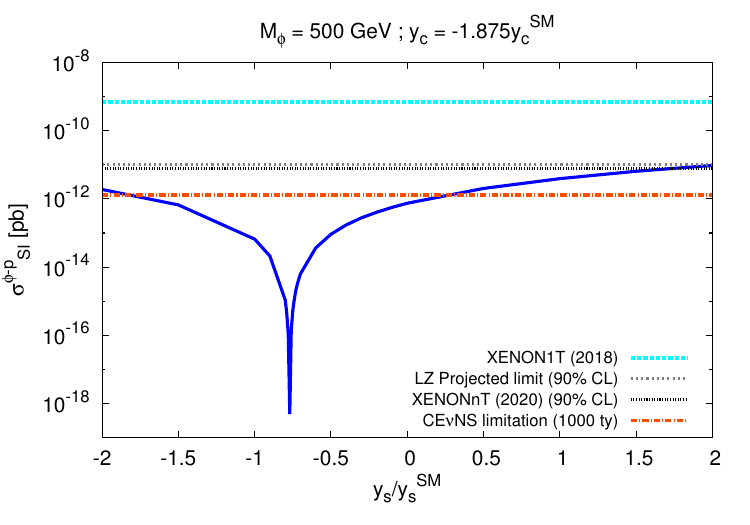}\label{fig:CrossC}}\quad
\subfloat[\quad\quad\quad(d)]{\includegraphics[scale=0.6]{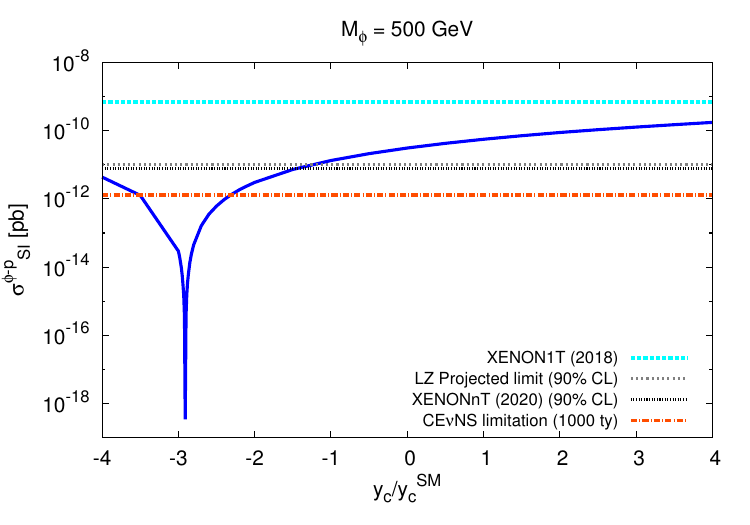}\label{fig:CrossD}}\\
\quad\subfloat[\quad\quad\quad(e)]{\includegraphics[scale=0.6]{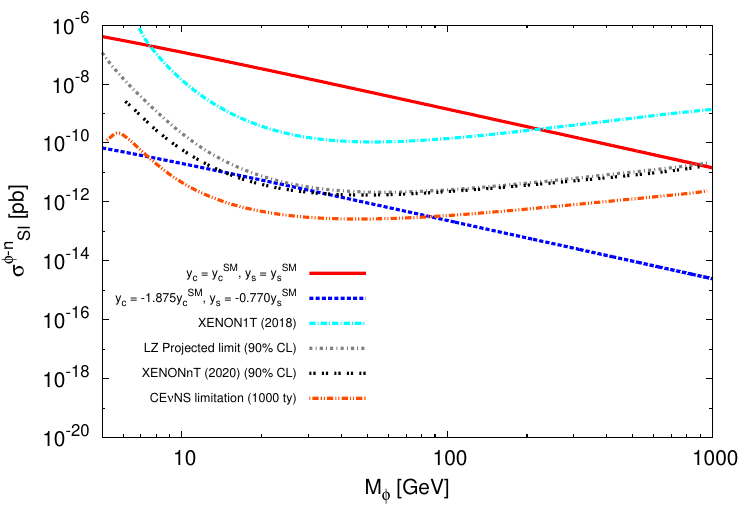}\label{fig:CrossE}}
\end{center}
\caption{Variation of the $\sigma_{\rm SI}^{\phi-p}$ with respect to $M_{\phi}$ and $y_s$ or $y_c$. In (a),
$y_c$ and $y_s$ are assumed to have non-SM values while
all other Yukawa couplings are fixed to their SM values and in (b), only $y_c$ can  assume non-SM values. In (c), the variation of $\sigma_{\rm SI}^{\phi-p}$ has been plotted as function of $y_s/y_s^{\rm SM}$ when only $y_c$ assumes non-SM value and in (d), only $y_c$ can have non-SM values, while $M_\phi=500$ GeV. In (e), we show the variation of $\sigma_{\rm SI}^{\phi-n}$ as a function of $M_{\phi}$ for the non-SM values of $y_c$ and $y_s$ for which $\sigma_{\rm SI}^{\phi-p}$ becomes minimum.}
\label{fig:Cross}
\end{figure}

The variation of $\Omega_{\rm \phi}h^2$ as a function of the DM mass is depicted in
Fig.~\ref{fig:Rplot}. The dependence of the relic density on the
variations of
charm and strange quark Yukawa couplings is negligible.
The blue dotted line in the figure represents the central value of the DM relic density obtained from the {\sc PLANCK} data. We see that, in addition to the resonance dip, this model is also able to satisfy the astrophysical data nicely for $M_{\rm \phi}\sim 100\pm 10$ GeV.  Even though the present DD bounds exclude a real singlet scalar DM in this region, the 
non-SM Yukawa couplings allow us to retain it, as we see from Fig.~\ref{fig:Cross}.\footnote{There can be specific realizations of WIMP models where a sub-TeV DM can be viable ~\cite{Bhattacharya:2017fid,Yepes:2018zkk}.} %

Fig.~\ref{fig:Cross} validates our
qualitative conclusions from
Sec.~\ref{sec:SIhiggs}. Here, we show the variation of
SI DD cross section as a function of 
$M_{\phi}$. We consider
two situations: (i)
both $y_c$ and $y_s$ can assume non-SM values while
all other Yukawa couplings are fixed to their SM values and (ii) only $y_c$ can assume non-SM values. One may simply
use Eq.~\eqref{eq:FN} to find $y_q$ such that the DM-nucleon SI scattering cross-section becomes
small enough to evade limits from the future-generation experiments. In the first
scenario, absolute
values of $y_c$ and
$y_s$ can be much closer to their SM values while in the second case, we find that a somewhat larger value, i.e., $y_c  \approx -2.9\,y_c^{\rm SM}$ would be necessary. Overall, the
negative values of light-quark Yukawa couplings help to accommodate the
DD bounds. For example, we see in Fig.~\ref{fig:CrossA} that there is an $\mathcal{O}(10^{-9})$ suppression between the SM line~(red) and the exact cancellation line~(blue). To illustrate, we can see that at $M_{\rm \phi}=100$ GeV, $\sigma^{\phi-p}_{\rm SI}(y_c=y_c^{\rm SM},\, y_s=y_s^{\rm SM})=1.384\times 10^{-9}$ pb, while $\sigma^{\phi-p}_{\rm SI}(y_c=-1.875~y_c^{\rm SM},\, y_s=-0.770~y_s^{\rm SM})=1.043\times 10^{-18}$ pb. At the backdrop of our non-SM Yukawa couplings, we have the VL quarks with mass about $2$~TeV. Thus, taking the NP scale $\Lambda\sim 2$ TeV, and with $m_{c,\, s}=m^{\rm SM}_{c,\, s}$, the said non-SM values of $y_c$ and $y_s$ can be achieved for $Y_H^c=1.76$ and $Y_H^s=0.07$, respectively.
Here the numerical value 
for $y_c$ is slightly away from that in Eq.~\eqref{eq:analyukawa},
due to the inclusion of QCD corrections.
Similar results are obtained in Fig.~\ref{fig:CrossB} as well. For illustrating the functional
dependence of $\sigma_{SI}^{\phi-p}$ with the 2nd  generation Yukawa couplings, we present Fig.~\ref{fig:CrossC} and Fig.~\ref{fig:CrossD}. In the
first plot, variation is shown over $y_s/y_s^{\rm SM}$ when $y_c=-1.875~y_c^{\rm SM}$ and in the 2nd plot, the
same variation is shown over $y_c/y_c^{\rm SM}$ when all other Yukawa couplings are SM-like, respectively. One may easily find out the correct numerical value to obtain the
desired cancellation in the DM-nucleon cross-section. Fig.~\ref{fig:CrossE} shows the variation of $\sigma_{SI}^{\phi-n}$ with respect to $M_\phi$, for 
the same set of $y_c$ and $y_s$ where $\lambda_p \rightarrow 0$, i.e., with $y_c=-1.875~y_c^{\rm SM}$ and $y_s=-0.770~y_s^{\rm SM}$, keeping all other quark Yukawa couplings fixed at their SM values. The blue line in Fig.~\ref{fig:CrossE} shows that even though  $\sigma_{SI}^{\phi-n}$ can't be vanishingly small for this set of parameter points,  the DM-neutron scattering cross section can be below the proposed direct detection bounds for $M_\phi\geq 50$ GeV. This, 
clearly, reflects 
isospin violation to a large extent, following from the constraint $\lambda_p \to 0$. 

However, the IVDM scenario can
be realized in a more general way and a large 
violation can be observed for moderate to large non-SM-like first-generation Yukawa couplings. Since
$f^{(N)}_{u,\, d}$ values are different for neutron and
proton, one may easily get a parameter space where
$f_p \neq f_n$. A particular useful scenario appears when ${f_n}/{f_p}$ takes a negative
value, as it then offers some significant cancellations in the 
DM-nucleus scattering cross section. Though a larger
value of $y_d$ or $y_u$ can easily lead to ${f_n}/{f_p}
>0 $, but negative values can only be achieved in a narrow domain which can be computed using Eq.~\eqref{eq:FN}. For example, considering only $y_d$ to assume non-SM values, the following range for the same coupling can be observed.
\begin{align}
    -\left(\frac{X^p}{f_d^p}\right)<\frac{y_d}{y_d^{\rm SM}}<-\left(\frac{X^n}{f_d^n}\right),
    \label{eq:XpXn}
\end{align}
where, 
\begin{align*}
  X^{(N)}=\sum_{q=u,s} f_q^{(N)}+\frac{2}{27}f^{(N)}_{G}\sum_{q=c,b,t} C_q\,,\quad\quad [N=p,n] . 
\end{align*} 
 Note that $X^p > X^n$ for $f_q^{(N)}$ shown in Eq.~\eqref{eq:ff}. For numerical estimation, we can consider a benchmark ratio, e.g., $f_n/f_p\approx-0.7$. This particular value has some importance to relax tensions between different results
 for low DM mass. One may easily check that, $y_d/y_d^{\rm SM}\approx-11.6$ can lead to the above ratio.
Finally, we note that in the limit of vanishing DM-nucleon couplings, the two-nucleon currents 
might become important~\cite{Korber:2017ery}.

%%%%%%%%%%%%%%%%%%%%%%%%%%%%%%%%%%%%%%%%%%%%%%%%%%%%%%%%%%%%%%%%%%%%%%%%%%%%%%%%
\section{Conclusion}
\label{sec:conclu}

\noindent
Direct searches of WIMPs as a form of dark matter have been underway for a long time. Due to the present and projected experimental sensitivities towards the spin-independent direct dark matter detection cross section, the available parameter spaces in simple H-portal dark matter models are significantly reduced or threatened to be ruled out. However, there exist a few small regions where the DM interactions with nucleons can be very tiny, thus escaping the ever impinging bounds from the DM searches. One such example is the so-called ``Blind spots'' where either the DM couplings with Higgs scalar vanish, or there is some destructive interference among diagrams involving different neutral scalars. In this paper, we have realized another route in the same direction, where the
Higgs boson couplings with the nucleons become vanishingly small. Apparently, such a requirement can be realized easily, if the light-quark Yukawa couplings are allowed to assume non-SM values in presence of some new physics; in particular, negative values --- a possibility allowed by the current experiments. Adopting a phenomenological effective theory perspective, we consider some higher dimensional operators or, in particular, dimension-6 operators involving the SM fields that let the light-quark Yukawa couplings to be negative without disturbing the respective quark masses. However, in a specific theory, precise cancellation of the DM-nucleon SI direct detection cross section may be realized only within a limited ranges of the parameters. Importantly, the new physics scale is bounded and for perturbative values of new couplings, can only be of the order of a few TeV. We also consider a specific realization of this with vector-like quarks with masses about $2$ TeV. In this set-up, we consider a real SM-singlet scalar as the DM candidate. In the absence of any discovery, generally, such a simple DM set up would be excluded completely for $M_{\rm DM} \le 1$ TeV from the projected sensitivity of the proposed LZ or XENONnT experiments. Here we observe that resultant SI DM-nucleon scattering cross section can be made vanishingly small for all values of DM mass. Needless to say, our observation would be unchanged for any other DM candidate in the Higgs-portal models or in the models where the Higgs produces the dominant contribution to the direct detection process, since our argument does not depend on the DM-DM-Higgs couplings and, hence, it can take $\mathcal O(1)$ values as well. 
Isospin-violation in the DM-nucleon scattering can also be realized for negative values of the first-generation light-quark Yukawa couplings. Usually, the exclusion limits are obtained assuming
isospin-conserving dark matter and hence can be much relaxed when the DM couples differently to protons and neutrons.
Probing such non-standard values of the Yukawa couplings of the first two generations of quarks is hard even for the HL-LHC, though there exist studies that aim to narrow down  the allowed values of the light-quark Yukawa couplings, even the negative values. For example, it has been argued that, at the LHC, it might be possible to pin the charm-quark Yukawa coupling within $[-0.6, 3]y_c^{\rm SM}$ with $3000$ fb$^{-1}$ of integrated luminosity in a largely model independent manner.  Hence, even though the future generation of dark matter search experiments based on the dark matter-nucleon scattering is blind to our proposal, it might be tested at the HL-LHC.

%%%%%%%%%%%%%%%%%%%%%%%%%%%%%%%%%%%%%%%%%%%%%%%%%%%%%%%%%%%%%%%%%%%%%%%%%%%%%%%%
\begin{acknowledgments}\noindent
We thank U. Chattopadhyay for some valuable comments. We also thank J. de Vries for pointing us towards the two-nucleon current contribution in DM-nucleus scattering. S. M. acknowledges support from the Science and Engineering Research Board (SERB), DST, India under Grant No. ECR/2017/000517. Our computations were supported in part by SAMKHYA: the High Performance Computing Facility provided by the Institute of Physics (IoP), Bhubaneswar, India. 

\end{acknowledgments}
%%%%%%%%%%%%%%%%%%%%%%%%%%%%%%%%%%%%%%%%%%%%%%%%%%%%%%%%%%%%%%%%%%%%%%%%%%%%%%%%

%%%%%%%%%%%%%%%%%% REFERENCES %%%%%%%%%%%%%%%%%%%%%%%%%%%%
\bigskip
\small \bibliography{dm_bsd}{}

\providecommand{\href}[2]{#2}\begingroup\raggedright\begin{thebibliography}{100}

\bibitem{Jungman:1995df}
G.~Jungman, M.~Kamionkowski and K.~Griest, \emph{{Supersymmetric dark matter}},
  \href{http://dx.doi.org/10.1016/0370-1573(95)00058-5}{\emph{Phys. Rept.} {\bf
  267} (1996) 195--373}, [\href{http://arxiv.org/abs/hep-ph/9506380}{{\tt
  hep-ph/9506380}}].

\bibitem{Bertone:2004pz}
G.~Bertone, D.~Hooper and J.~Silk, \emph{{Particle dark matter: Evidence,
  candidates and constraints}},
  \href{http://dx.doi.org/10.1016/j.physrep.2004.08.031}{\emph{Phys. Rept.}
  {\bf 405} (2005) 279--390}, [\href{http://arxiv.org/abs/hep-ph/0404175}{{\tt
  hep-ph/0404175}}].

\bibitem{Aprile:2018dbl}
{\scshape XENON} collaboration, E.~Aprile et~al., \emph{{Dark Matter Search
  Results from a One Ton-Year Exposure of XENON1T}},
  \href{http://dx.doi.org/10.1103/PhysRevLett.121.111302}{\emph{Phys. Rev.
  Lett.} {\bf 121} (2018) 111302}, [\href{http://arxiv.org/abs/1805.12562}{{\tt
  1805.12562}}].

\bibitem{Akerib:2017kat}
{\scshape LUX} collaboration, D.~S. Akerib et~al., \emph{{Limits on
  spin-dependent WIMP-nucleon cross section obtained from the complete LUX
  exposure}},
  \href{http://dx.doi.org/10.1103/PhysRevLett.118.251302}{\emph{Phys. Rev.
  Lett.} {\bf 118} (2017) 251302}, [\href{http://arxiv.org/abs/1705.03380}{{\tt
  1705.03380}}].

\bibitem{Cui:2017nnn}
{\scshape PandaX-II} collaboration, X.~Cui et~al., \emph{{Dark Matter Results
  From 54-Ton-Day Exposure of PandaX-II Experiment}},
  \href{http://dx.doi.org/10.1103/PhysRevLett.119.181302}{\emph{Phys. Rev.
  Lett.} {\bf 119} (2017) 181302}, [\href{http://arxiv.org/abs/1708.06917}{{\tt
  1708.06917}}].

\bibitem{PhysRevD.101.052008}
{\scshape SuperCDMS Collaboration} collaboration, T.~Aralis et~al.,
  \emph{Constraints on dark photons and axionlike particles from the supercdms
  soudan experiment},
  \href{http://dx.doi.org/10.1103/PhysRevD.101.052008}{\emph{Phys. Rev. D} {\bf
  101} (Mar, 2020) 052008}.

\bibitem{PhysRevD.99.062001}
{\scshape SuperCDMS Collaboration} collaboration, R.~Agnese et~al.,
  \emph{Search for low-mass dark matter with cdmslite using a profile
  likelihood fit},
  \href{http://dx.doi.org/10.1103/PhysRevD.99.062001}{\emph{Phys. Rev. D} {\bf
  99} (Mar, 2019) 062001}.

\bibitem{Undagoitia:2015gya}
T.~Marrod\'an~Undagoitia and L.~Rauch, \emph{{Dark matter direct-detection
  experiments}},
  \href{http://dx.doi.org/10.1088/0954-3899/43/1/013001}{\emph{J. Phys. G} {\bf
  43} (2016) 013001}, [\href{http://arxiv.org/abs/1509.08767}{{\tt
  1509.08767}}].

\bibitem{Roszkowski:2017nbc}
L.~Roszkowski, E.~M. Sessolo and S.~Trojanowski, \emph{{WIMP dark matter
  candidates and searches-current status and future prospects}},
  \href{http://dx.doi.org/10.1088/1361-6633/aab913}{\emph{Rept. Prog. Phys.}
  {\bf 81} (2018) 066201}, [\href{http://arxiv.org/abs/1707.06277}{{\tt
  1707.06277}}].

\bibitem{Schumann:2019eaa}
M.~Schumann, \emph{{Direct Detection of WIMP Dark Matter: Concepts and
  Status}}, \href{http://dx.doi.org/10.1088/1361-6471/ab2ea5}{\emph{J. Phys. G}
  {\bf 46} (2019) 103003}, [\href{http://arxiv.org/abs/1903.03026}{{\tt
  1903.03026}}].

\bibitem{Akerib:2018lyp}
{\scshape LUX-ZEPLIN} collaboration, D.~S. Akerib et~al., \emph{{Projected WIMP
  Sensitivity of the LUX-ZEPLIN (LZ) Dark Matter Experiment}},
  \href{http://arxiv.org/abs/1802.06039}{{\tt 1802.06039}}.

\bibitem{Aprile:2020vtw}
{\scshape XENON} collaboration, E.~Aprile et~al., \emph{{Projected WIMP
  Sensitivity of the XENONnT Dark Matter Experiment}},
  \href{http://arxiv.org/abs/2007.08796}{{\tt 2007.08796}}.

\bibitem{Arcadi:2014lta}
G.~Arcadi, Y.~Mambrini and F.~Richard, \emph{{Z-portal dark matter}},
  \href{http://dx.doi.org/10.1088/1475-7516/2015/03/018}{\emph{JCAP} {\bf 1503}
  (2015) 018}, [\href{http://arxiv.org/abs/1411.2985}{{\tt 1411.2985}}].

\bibitem{Hamaguchi:2015rxa}
K.~Hamaguchi and K.~Ishikawa, \emph{{Prospects for Higgs- and Z-resonant
  Neutralino Dark Matter}},
  \href{http://dx.doi.org/10.1103/PhysRevD.93.055009}{\emph{Phys. Rev.} {\bf
  D93} (2016) 055009}, [\href{http://arxiv.org/abs/1510.05378}{{\tt
  1510.05378}}].

\bibitem{Escudero:2016gzx}
M.~Escudero, A.~Berlin, D.~Hooper and M.-X. Lin, \emph{{Toward (Finally!)
  Ruling Out Z and Higgs Mediated Dark Matter Models}},
  \href{http://dx.doi.org/10.1088/1475-7516/2016/12/029}{\emph{JCAP} {\bf 1612}
  (2016) 029}, [\href{http://arxiv.org/abs/1609.09079}{{\tt 1609.09079}}].

\bibitem{Kearney:2016rng}
J.~Kearney, N.~Orlofsky and A.~Pierce, \emph{{$Z$ boson mediated dark matter
  beyond the effective theory}},
  \href{http://dx.doi.org/10.1103/PhysRevD.95.035020}{\emph{Phys. Rev.} {\bf
  D95} (2017) 035020}, [\href{http://arxiv.org/abs/1611.05048}{{\tt
  1611.05048}}].

\bibitem{Balazs:2017ple}
J.~Ellis, A.~Fowlie, L.~Marzola and M.~Raidal, \emph{{Statistical Analyses of
  Higgs- and Z-Portal Dark Matter Models}},
  \href{http://dx.doi.org/10.1103/PhysRevD.97.115014}{\emph{Phys. Rev.} {\bf
  D97} (2018) 115014}, [\href{http://arxiv.org/abs/1711.09912}{{\tt
  1711.09912}}].

\bibitem{Barman:2019mlj}
B.~Barman, T.~Bhanja, D.~Das and D.~Maity, \emph{{Minimal model of torsion
  mediated dark matter}},
  \href{http://dx.doi.org/10.1103/PhysRevD.101.075017}{\emph{Phys. Rev. D} {\bf
  101} (2020) 075017}, [\href{http://arxiv.org/abs/1912.09249}{{\tt
  1912.09249}}].

\bibitem{Silveira:1985rk}
V.~Silveira and A.~Zee, \emph{{SCALAR PHANTOMS}},
  \href{http://dx.doi.org/10.1016/0370-2693(85)90624-0}{\emph{Phys. Lett. B}
  {\bf 161} (1985) 136--140}.

\bibitem{McDonald:1993ex}
J.~McDonald, \emph{{Gauge singlet scalars as cold dark matter}},
  \href{http://dx.doi.org/10.1103/PhysRevD.50.3637}{\emph{Phys. Rev. D} {\bf
  50} (1994) 3637--3649}, [\href{http://arxiv.org/abs/hep-ph/0702143}{{\tt
  hep-ph/0702143}}].

\bibitem{Burgess:2000yq}
C.~Burgess, M.~Pospelov and T.~ter Veldhuis, \emph{{The Minimal model of
  nonbaryonic dark matter: A Singlet scalar}},
  \href{http://dx.doi.org/10.1016/S0550-3213(01)00513-2}{\emph{Nucl. Phys. B}
  {\bf 619} (2001) 709--728}, [\href{http://arxiv.org/abs/hep-ph/0011335}{{\tt
  hep-ph/0011335}}].

\bibitem{Lebedev:2011iq}
O.~Lebedev, H.~M. Lee and Y.~Mambrini, \emph{{Vector Higgs-portal dark matter
  and the invisible Higgs}},
  \href{http://dx.doi.org/10.1016/j.physletb.2012.01.029}{\emph{Phys. Lett.}
  {\bf B707} (2012) 570--576}, [\href{http://arxiv.org/abs/1111.4482}{{\tt
  1111.4482}}].

\bibitem{Djouadi:2011aa}
A.~Djouadi, O.~Lebedev, Y.~Mambrini and J.~Quevillon, \emph{{Implications of
  LHC searches for Higgs--portal dark matter}},
  \href{http://dx.doi.org/10.1016/j.physletb.2012.01.062}{\emph{Phys. Lett.}
  {\bf B709} (2012) 65--69}, [\href{http://arxiv.org/abs/1112.3299}{{\tt
  1112.3299}}].

\bibitem{Djouadi:2012zc}
A.~Djouadi, A.~Falkowski, Y.~Mambrini and J.~Quevillon, \emph{{Direct Detection
  of Higgs-Portal Dark Matter at the LHC}},
  \href{http://dx.doi.org/10.1140/epjc/s10052-013-2455-1}{\emph{Eur. Phys. J.
  C} {\bf 73} (2013) 2455}, [\href{http://arxiv.org/abs/1205.3169}{{\tt
  1205.3169}}].

\bibitem{Cline:2013gha}
J.~M. Cline, K.~Kainulainen, P.~Scott and C.~Weniger, \emph{{Update on scalar
  singlet dark matter}},
  \href{http://dx.doi.org/10.1103/PhysRevD.88.055025}{\emph{Phys. Rev. D} {\bf
  88} (2013) 055025}, [\href{http://arxiv.org/abs/1306.4710}{{\tt 1306.4710}}].
  [Erratum: Phys.Rev.D 92, 039906 (2015)].

\bibitem{Gross:2015cwa}
C.~Gross, O.~Lebedev and Y.~Mambrini, \emph{{Non-Abelian gauge fields as dark
  matter}}, \href{http://dx.doi.org/10.1007/JHEP08(2015)158}{\emph{JHEP} {\bf
  08} (2015) 158}, [\href{http://arxiv.org/abs/1505.07480}{{\tt 1505.07480}}].

\bibitem{Casas:2017jjg}
J.~A. Casas, D.~G. Cerdeño, J.~M. Moreno and J.~Quilis, \emph{{Reopening the
  Higgs portal for single scalar dark matter}},
  \href{http://dx.doi.org/10.1007/JHEP05(2017)036}{\emph{JHEP} {\bf 05} (2017)
  036}, [\href{http://arxiv.org/abs/1701.08134}{{\tt 1701.08134}}].

\bibitem{Hoferichter:2017olk}
M.~Hoferichter, P.~Klos, J.~Men\'endez and A.~Schwenk, \emph{{Improved limits
  for Higgs-portal dark matter from LHC searches}},
  \href{http://dx.doi.org/10.1103/PhysRevLett.119.181803}{\emph{Phys. Rev.
  Lett.} {\bf 119} (2017) 181803}, [\href{http://arxiv.org/abs/1708.02245}{{\tt
  1708.02245}}].

\bibitem{Arcadi:2019lka}
G.~Arcadi, A.~Djouadi and M.~Raidal, \emph{{Dark Matter through the Higgs
  portal}},  \href{http://arxiv.org/abs/1903.03616}{{\tt 1903.03616}}.

\bibitem{Mambrini:2010dq}
Y.~Mambrini, \emph{{The Kinetic dark-mixing in the light of CoGENT and
  XENON100}},
  \href{http://dx.doi.org/10.1088/1475-7516/2010/09/022}{\emph{JCAP} {\bf 1009}
  (2010) 022}, [\href{http://arxiv.org/abs/1006.3318}{{\tt 1006.3318}}].

\bibitem{Dudas:2013sia}
E.~Dudas, L.~Heurtier, Y.~Mambrini and B.~Zaldivar, \emph{{Extra U(1),
  effective operators, anomalies and dark matter}},
  \href{http://dx.doi.org/10.1007/JHEP11(2013)083}{\emph{JHEP} {\bf 11} (2013)
  083}, [\href{http://arxiv.org/abs/1307.0005}{{\tt 1307.0005}}].

\bibitem{Alves:2013tqa}
A.~Alves, S.~Profumo and F.~S. Queiroz, \emph{{The dark $Z^{'}$ portal: direct,
  indirect and collider searches}},
  \href{http://dx.doi.org/10.1007/JHEP04(2014)063}{\emph{JHEP} {\bf 04} (2014)
  063}, [\href{http://arxiv.org/abs/1312.5281}{{\tt 1312.5281}}].

\bibitem{Lebedev:2014bba}
O.~Lebedev and Y.~Mambrini, \emph{{Axial dark matter: The case for an invisible
  $Z'$}}, \href{http://dx.doi.org/10.1016/j.physletb.2014.05.025}{\emph{Phys.
  Lett. B} {\bf 734} (2014) 350--353},
  [\href{http://arxiv.org/abs/1403.4837}{{\tt 1403.4837}}].

\bibitem{Hooper:2014fda}
D.~Hooper, \emph{{$Z^\prime$ Mediated Dark Matter Models for the Galactic
  Center Gamma-Ray Excess}},
  \href{http://dx.doi.org/10.1103/PhysRevD.91.035025}{\emph{Phys. Rev. D} {\bf
  91} (2015) 035025}, [\href{http://arxiv.org/abs/1411.4079}{{\tt 1411.4079}}].

\bibitem{Alves:2015pea}
A.~Alves, A.~Berlin, S.~Profumo and F.~S. Queiroz, \emph{{Dark Matter
  Complementarity and the Z$^\prime$ Portal}},
  \href{http://dx.doi.org/10.1103/PhysRevD.92.083004}{\emph{Phys. Rev.} {\bf
  D92} (2015) 083004}, [\href{http://arxiv.org/abs/1501.03490}{{\tt
  1501.03490}}].

\bibitem{Alves:2015mua}
A.~Alves, A.~Berlin, S.~Profumo and F.~S. Queiroz, \emph{{Dirac-fermionic dark
  matter in U(1)$_{X}$ models}},
  \href{http://dx.doi.org/10.1007/JHEP10(2015)076}{\emph{JHEP} {\bf 10} (2015)
  076}, [\href{http://arxiv.org/abs/1506.06767}{{\tt 1506.06767}}].

\bibitem{Allanach:2015gkd}
B.~Allanach, F.~S. Queiroz, A.~Strumia and S.~Sun, \emph{{$Z^{\prime}$ models
  for the LHCb and $g-2$ muon anomalies}},
  \href{http://dx.doi.org/10.1103/PhysRevD.93.055045,
  10.1103/PhysRevD.95.119902}{\emph{Phys. Rev.} {\bf D93} (2016) 055045},
  [\href{http://arxiv.org/abs/1511.07447}{{\tt 1511.07447}}]. [Erratum: Phys.
  Rev.D95,no.11,119902(2017)].

\bibitem{Alves:2016cqf}
A.~Alves, G.~Arcadi, Y.~Mambrini, S.~Profumo and F.~S. Queiroz, \emph{{Augury
  of darkness: the low-mass dark Z$^{\prime}$ portal}},
  \href{http://dx.doi.org/10.1007/JHEP04(2017)164}{\emph{JHEP} {\bf 04} (2017)
  164}, [\href{http://arxiv.org/abs/1612.07282}{{\tt 1612.07282}}].

\bibitem{Berlin:2015wwa}
A.~Berlin, S.~Gori, T.~Lin and L.-T. Wang, \emph{{Pseudoscalar Portal Dark
  Matter}}, \href{http://dx.doi.org/10.1103/PhysRevD.92.015005}{\emph{Phys.
  Rev.} {\bf D92} (2015) 015005}, [\href{http://arxiv.org/abs/1502.06000}{{\tt
  1502.06000}}].

\bibitem{Baek:2017vzd}
S.~Baek, P.~Ko and J.~Li, \emph{{Minimal renormalizable simplified dark matter
  model with a pseudoscalar mediator}},
  \href{http://dx.doi.org/10.1103/PhysRevD.95.075011}{\emph{Phys. Rev. D} {\bf
  95} (2017) 075011}, [\href{http://arxiv.org/abs/1701.04131}{{\tt
  1701.04131}}].

\bibitem{Bauer:2017fsw}
M.~Bauer, M.~Klassen and V.~Tenorth, \emph{{Universal properties of
  pseudoscalar mediators in dark matter extensions of 2HDMs}},
  \href{http://dx.doi.org/10.1007/JHEP07(2018)107}{\emph{JHEP} {\bf 07} (2018)
  107}, [\href{http://arxiv.org/abs/1712.06597}{{\tt 1712.06597}}].

\bibitem{Arcadi:2017kky}
G.~Arcadi, M.~Dutra, P.~Ghosh, M.~Lindner, Y.~Mambrini, M.~Pierre et~al.,
  \emph{{The waning of the WIMP? A review of models, searches, and
  constraints}},
  \href{http://dx.doi.org/10.1140/epjc/s10052-018-5662-y}{\emph{Eur. Phys. J.}
  {\bf C78} (2018) 203}, [\href{http://arxiv.org/abs/1703.07364}{{\tt
  1703.07364}}].

\bibitem{Drees:1992am}
M.~Drees and M.~M. Nojiri, \emph{{The Neutralino relic density in minimal $N=1$
  supergravity}}, \href{http://dx.doi.org/10.1103/PhysRevD.47.376}{\emph{Phys.
  Rev.} {\bf D47} (1993) 376--408},
  [\href{http://arxiv.org/abs/hep-ph/9207234}{{\tt hep-ph/9207234}}].

\bibitem{Nath:1992ty}
P.~Nath and R.~L. Arnowitt, \emph{{Predictions in SU(5) supergravity grand
  unification with proton stability and relic density constraints}},
  \href{http://dx.doi.org/10.1103/PhysRevLett.70.3696}{\emph{Phys. Rev. Lett.}
  {\bf 70} (1993) 3696--3699}, [\href{http://arxiv.org/abs/hep-ph/9302318}{{\tt
  hep-ph/9302318}}].

\bibitem{Baer:1997ai}
H.~Baer and M.~Brhlik, \emph{{Neutralino dark matter in minimal supergravity:
  Direct detection versus collider searches}},
  \href{http://dx.doi.org/10.1103/PhysRevD.57.567}{\emph{Phys. Rev.} {\bf D57}
  (1998) 567--577}, [\href{http://arxiv.org/abs/hep-ph/9706509}{{\tt
  hep-ph/9706509}}].

\bibitem{Ellis:2003cw}
J.~R. Ellis, K.~A. Olive, Y.~Santoso and V.~C. Spanos, \emph{{Supersymmetric
  dark matter in light of WMAP}},
  \href{http://dx.doi.org/10.1016/S0370-2693(03)00765-2}{\emph{Phys. Lett. B}
  {\bf 565} (2003) 176--182}, [\href{http://arxiv.org/abs/hep-ph/0303043}{{\tt
  hep-ph/0303043}}].

\bibitem{Chattopadhyay:2008hk}
U.~Chattopadhyay and D.~Das, \emph{{Higgs funnel region of SUSY dark matter for
  small tan beta, RG effects on pseudoscalar Higgs boson with scalar mass
  non-universality}},
  \href{http://dx.doi.org/10.1103/PhysRevD.79.035007}{\emph{Phys. Rev.} {\bf
  D79} (2009) 035007}, [\href{http://arxiv.org/abs/0809.4065}{{\tt
  0809.4065}}].

\bibitem{Chattopadhyay:2010vp}
U.~Chattopadhyay, D.~Das, D.~K. Ghosh and M.~Maity, \emph{{Probing the light
  Higgs pole resonance annihilation of dark matter in the light of XENON100 and
  CDMS-II observations}},
  \href{http://dx.doi.org/10.1103/PhysRevD.82.075013}{\emph{Phys. Rev.} {\bf
  D82} (2010) 075013}, [\href{http://arxiv.org/abs/1006.3045}{{\tt
  1006.3045}}].

\bibitem{Das:2010kb}
D.~Das, A.~Goudelis and Y.~Mambrini, \emph{{Exploring SUSY light Higgs boson
  scenarios via dark matter experiments}},
  \href{http://dx.doi.org/10.1088/1475-7516/2010/12/018}{\emph{JCAP} {\bf 1012}
  (2010) 018}, [\href{http://arxiv.org/abs/1007.4812}{{\tt 1007.4812}}].

\bibitem{Chatterjee:2014bva}
A.~Chatterjee, D.~Das, B.~Mukhopadhyaya and S.~K. Rai, \emph{{Right Sneutrino
  Dark Matter and a Monochromatic Photon Line}},
  \href{http://dx.doi.org/10.1088/1475-7516/2014/07/023}{\emph{JCAP} {\bf 07}
  (2014) 023}, [\href{http://arxiv.org/abs/1401.2527}{{\tt 1401.2527}}].

\bibitem{Feng:2010gw}
J.~L. Feng, \emph{{Dark Matter Candidates from Particle Physics and Methods of
  Detection}},
  \href{http://dx.doi.org/10.1146/annurev-astro-082708-101659}{\emph{Ann. Rev.
  Astron. Astrophys.} {\bf 48} (2010) 495--545},
  [\href{http://arxiv.org/abs/1003.0904}{{\tt 1003.0904}}].

\bibitem{He:2008qm}
X.-G. He, T.~Li, X.-Q. Li, J.~Tandean and H.-C. Tsai, \emph{{Constraints on
  Scalar Dark Matter from Direct Experimental Searches}},
  \href{http://dx.doi.org/10.1103/PhysRevD.79.023521}{\emph{Phys. Rev. D} {\bf
  79} (2009) 023521}, [\href{http://arxiv.org/abs/0811.0658}{{\tt 0811.0658}}].

\bibitem{He:2011gc}
X.-G. He, B.~Ren and J.~Tandean, \emph{{Hints of Standard Model Higgs Boson at
  the LHC and Light Dark Matter Searches}},
  \href{http://dx.doi.org/10.1103/PhysRevD.85.093019}{\emph{Phys. Rev. D} {\bf
  85} (2012) 093019}, [\href{http://arxiv.org/abs/1112.6364}{{\tt 1112.6364}}].

\bibitem{Cheung:2012qy}
C.~Cheung, L.~J. Hall, D.~Pinner and J.~T. Ruderman, \emph{{Prospects and Blind
  Spots for Neutralino Dark Matter}},
  \href{http://dx.doi.org/10.1007/JHEP05(2013)100}{\emph{JHEP} {\bf 05} (2013)
  100}, [\href{http://arxiv.org/abs/1211.4873}{{\tt 1211.4873}}].

\bibitem{Chang:2017gla}
C.-F. Chang, X.-G. He and J.~Tandean, \emph{{Two-Higgs-Doublet-Portal
  Dark-Matter Models in Light of Direct Search and LHC Data}},
  \href{http://dx.doi.org/10.1007/JHEP04(2017)107}{\emph{JHEP} {\bf 04} (2017)
  107}, [\href{http://arxiv.org/abs/1702.02924}{{\tt 1702.02924}}].

\bibitem{Huang:2014xua}
P.~Huang and C.~E.~M. Wagner, \emph{{Blind Spots for neutralino Dark Matter in
  the MSSM with an intermediate $m_A$}},
  \href{http://dx.doi.org/10.1103/PhysRevD.90.015018}{\emph{Phys. Rev. D} {\bf
  90} (2014) 015018}, [\href{http://arxiv.org/abs/1404.0392}{{\tt 1404.0392}}].

\bibitem{Badziak:2015exr}
M.~Badziak, M.~Olechowski and P.~Szczerbiak, \emph{{Blind spots for neutralino
  dark matter in the NMSSM}},
  \href{http://dx.doi.org/10.1007/JHEP03(2016)179}{\emph{JHEP} {\bf 03} (2016)
  179}, [\href{http://arxiv.org/abs/1512.02472}{{\tt 1512.02472}}].

\bibitem{Crivellin:2015bva}
A.~Crivellin, M.~Hoferichter, M.~Procura and L.~C. Tunstall, \emph{{Light
  stops, blind spots, and isospin violation in the MSSM}},
  \href{http://dx.doi.org/10.1007/JHEP07(2015)129}{\emph{JHEP} {\bf 07} (2015)
  129}, [\href{http://arxiv.org/abs/1503.03478}{{\tt 1503.03478}}].

\bibitem{Han:2016qtc}
T.~Han, F.~Kling, S.~Su and Y.~Wu, \emph{{Unblinding the dark matter blind
  spots}}, \href{http://dx.doi.org/10.1007/JHEP02(2017)057}{\emph{JHEP} {\bf
  02} (2017) 057}, [\href{http://arxiv.org/abs/1612.02387}{{\tt 1612.02387}}].

\bibitem{Alanne:2017oqj}
T.~Alanne and F.~Goertz, \emph{{Extended Dark Matter EFT}},
  \href{http://dx.doi.org/10.1140/epjc/s10052-020-7999-2}{\emph{Eur. Phys. J.
  C} {\bf 80} (2020) 446}, [\href{http://arxiv.org/abs/1712.07626}{{\tt
  1712.07626}}].

\bibitem{Altmannshofer:2019wjb}
W.~Altmannshofer, B.~Maddock and S.~Profumo, \emph{{Doubly Blind Spots in
  Scalar Dark Matter Models}},
  \href{http://dx.doi.org/10.1103/PhysRevD.100.055033}{\emph{Phys. Rev. D} {\bf
  100} (2019) 055033}, [\href{http://arxiv.org/abs/1907.01726}{{\tt
  1907.01726}}].

\bibitem{Alanne:2020xcb}
T.~Alanne, G.~Arcadi, F.~Goertz, V.~Tenorth and S.~Vogl,
  \emph{{Model-independent constraints with extended dark matter EFT}},
  \href{http://dx.doi.org/10.1007/JHEP10(2020)172}{\emph{JHEP} {\bf 10} (2020)
  172}, [\href{http://arxiv.org/abs/2006.07174}{{\tt 2006.07174}}].

\bibitem{Gross:2017dan}
C.~Gross, O.~Lebedev and T.~Toma, \emph{{Cancellation Mechanism for Dark-Matter
  Nucleon Interaction}},
  \href{http://dx.doi.org/10.1103/PhysRevLett.119.191801}{\emph{Phys. Rev.
  Lett.} {\bf 119} (2017) 191801}, [\href{http://arxiv.org/abs/1708.02253}{{\tt
  1708.02253}}].

\bibitem{Balkin:2018tma}
R.~Balkin, M.~Ruhdorfer, E.~Salvioni and A.~Weiler, \emph{{Dark matter shifts
  away from direct detection}},
  \href{http://dx.doi.org/10.1088/1475-7516/2018/11/050}{\emph{JCAP} {\bf 11}
  (2018) 050}, [\href{http://arxiv.org/abs/1809.09106}{{\tt 1809.09106}}].

\bibitem{Kurylov:2003ra}
A.~Kurylov and M.~Kamionkowski, \emph{{Generalized analysis of weakly
  interacting massive particle searches}},
  \href{http://dx.doi.org/10.1103/PhysRevD.69.063503}{\emph{Phys. Rev. D} {\bf
  69} (2004) 063503}, [\href{http://arxiv.org/abs/hep-ph/0307185}{{\tt
  hep-ph/0307185}}].

\bibitem{Giuliani:2005my}
F.~Giuliani, \emph{{Are direct search experiments sensitive to all
  spin-independent WIMP candidates?}},
  \href{http://dx.doi.org/10.1103/PhysRevLett.95.101301}{\emph{Phys. Rev.
  Lett.} {\bf 95} (2005) 101301},
  [\href{http://arxiv.org/abs/hep-ph/0504157}{{\tt hep-ph/0504157}}].

\bibitem{Chang:2010yk}
S.~Chang, J.~Liu, A.~Pierce, N.~Weiner and I.~Yavin, \emph{{CoGeNT
  Interpretations}},
  \href{http://dx.doi.org/10.1088/1475-7516/2010/08/018}{\emph{JCAP} {\bf 08}
  (2010) 018}, [\href{http://arxiv.org/abs/1004.0697}{{\tt 1004.0697}}].

\bibitem{Kang:2010mh}
Z.~Kang, T.~Li, T.~Liu, C.~Tong and J.~M. Yang, \emph{{Light Dark Matter from
  the $U(1)_X$ Sector in the NMSSM with Gauge Mediation}},
  \href{http://dx.doi.org/10.1088/1475-7516/2011/01/028}{\emph{JCAP} {\bf 01}
  (2011) 028}, [\href{http://arxiv.org/abs/1008.5243}{{\tt 1008.5243}}].

\bibitem{Feng:2011vu}
J.~L. Feng, J.~Kumar, D.~Marfatia and D.~Sanford, \emph{{Isospin-Violating Dark
  Matter}}, \href{http://dx.doi.org/10.1016/j.physletb.2011.07.083}{\emph{Phys.
  Lett. B} {\bf 703} (2011) 124--127},
  [\href{http://arxiv.org/abs/1102.4331}{{\tt 1102.4331}}].

\bibitem{Feng:2013vod}
J.~L. Feng, J.~Kumar and D.~Sanford, \emph{{Xenophobic Dark Matter}},
  \href{http://dx.doi.org/10.1103/PhysRevD.88.015021}{\emph{Phys. Rev. D} {\bf
  88} (2013) 015021}, [\href{http://arxiv.org/abs/1306.2315}{{\tt 1306.2315}}].

\bibitem{Yaguna:2016bga}
C.~E. Yaguna, \emph{{Isospin-violating dark matter in the light of recent
  data}}, \href{http://dx.doi.org/10.1103/PhysRevD.95.055015}{\emph{Phys. Rev.
  D} {\bf 95} (2017) 055015}, [\href{http://arxiv.org/abs/1610.08683}{{\tt
  1610.08683}}].

\bibitem{Bishara:2015cha}
F.~Bishara, J.~Brod, P.~Uttayarat and J.~Zupan, \emph{{Nonstandard Yukawa
  Couplings and Higgs Portal Dark Matter}},
  \href{http://dx.doi.org/10.1007/JHEP01(2016)010}{\emph{JHEP} {\bf 01} (2016)
  010}, [\href{http://arxiv.org/abs/1504.04022}{{\tt 1504.04022}}].

\bibitem{Buchmuller:1985jz}
W.~Buchmuller and D.~Wyler, \emph{{Effective Lagrangian Analysis of New
  Interactions and Flavor Conservation}},
  \href{http://dx.doi.org/10.1016/0550-3213(86)90262-2}{\emph{Nucl. Phys. B}
  {\bf 268} (1986) 621--653}.

\bibitem{Grzadkowski:2010es}
B.~Grzadkowski, M.~Iskrzynski, M.~Misiak and J.~Rosiek, \emph{{Dimension-Six
  Terms in the Standard Model Lagrangian}},
  \href{http://dx.doi.org/10.1007/JHEP10(2010)085}{\emph{JHEP} {\bf 10} (2010)
  085}, [\href{http://arxiv.org/abs/1008.4884}{{\tt 1008.4884}}].

\bibitem{Bhaskar:2020kdr}
A.~Bhaskar, D.~Das, B.~De and S.~Mitra, \emph{{Enhancing scalar productions
  with leptoquarks at the LHC}},
  \href{http://dx.doi.org/10.1103/PhysRevD.102.035002}{\emph{Phys. Rev. D} {\bf
  102} (2020) 035002}, [\href{http://arxiv.org/abs/2002.12571}{{\tt
  2002.12571}}].

\bibitem{Aad:2019mbh}
{\scshape ATLAS} collaboration, G.~Aad et~al., \emph{{Combined measurements of
  Higgs boson production and decay using up to $80$ fb$^{-1}$ of proton-proton
  collision data at $\sqrt{s}=$ 13 TeV collected with the ATLAS experiment}},
  \href{http://dx.doi.org/10.1103/PhysRevD.101.012002}{\emph{Phys. Rev. D} {\bf
  101} (2020) 012002}, [\href{http://arxiv.org/abs/1909.02845}{{\tt
  1909.02845}}].

\bibitem{deBlas:2019rxi}
J.~de~Blas et~al., \emph{{Higgs Boson Studies at Future Particle Colliders}},
  \href{http://dx.doi.org/10.1007/JHEP01(2020)139}{\emph{JHEP} {\bf 01} (2020)
  139}, [\href{http://arxiv.org/abs/1905.03764}{{\tt 1905.03764}}].

\bibitem{Falkowski:2020znk}
A.~Falkowski, S.~Ganguly, P.~Gras, J.~M. No, K.~Tobioka, N.~Vignaroli et~al.,
  \emph{{Light quark Yukawas in triboson final states}},
  \href{http://arxiv.org/abs/2011.09551}{{\tt 2011.09551}}.

\bibitem{Hedri:2013wea}
S.~El~Hedri, P.~J. Fox and J.~G. Wacker, \emph{{Exploring the Dark Side of the
  Top Yukawa}},  \href{http://arxiv.org/abs/1311.6488}{{\tt 1311.6488}}.

\bibitem{Egana-Ugrinovic:2018znw}
D.~Egana-Ugrinovic, S.~Homiller and P.~Meade, \emph{{Aligned and Spontaneous
  Flavor Violation}},
  \href{http://dx.doi.org/10.1103/PhysRevLett.123.031802}{\emph{Phys. Rev.
  Lett.} {\bf 123} (2019) 031802}, [\href{http://arxiv.org/abs/1811.00017}{{\tt
  1811.00017}}].

\bibitem{Egana-Ugrinovic:2019dqu}
D.~Egana-Ugrinovic, S.~Homiller and P.~R. Meade, \emph{{Higgs bosons with large
  couplings to light quarks}},
  \href{http://dx.doi.org/10.1103/PhysRevD.100.115041}{\emph{Phys. Rev. D} {\bf
  100} (2019) 115041}, [\href{http://arxiv.org/abs/1908.11376}{{\tt
  1908.11376}}].

\bibitem{Bar-Shalom:2018rjs}
S.~Bar-Shalom and A.~Soni, \emph{{Universally enhanced light-quarks Yukawa
  couplings paradigm}},
  \href{http://dx.doi.org/10.1103/PhysRevD.98.055001}{\emph{Phys. Rev.} {\bf
  D98} (2018) 055001}, [\href{http://arxiv.org/abs/1804.02400}{{\tt
  1804.02400}}].

\bibitem{Bishara:2016jga}
F.~Bishara, U.~Haisch, P.~F. Monni and E.~Re, \emph{{Constraining Light-Quark
  Yukawa Couplings from Higgs Distributions}},
  \href{http://dx.doi.org/10.1103/PhysRevLett.118.121801}{\emph{Phys. Rev.
  Lett.} {\bf 118} (2017) 121801}, [\href{http://arxiv.org/abs/1606.09253}{{\tt
  1606.09253}}].

\bibitem{Bonner:2016sdg}
G.~Bonner and H.~E. Logan, \emph{{Constraining the Higgs couplings to up and
  down quarks using production kinematics at the CERN Large Hadron Collider}},
  \href{http://arxiv.org/abs/1608.04376}{{\tt 1608.04376}}.

\bibitem{SHIFMAN1978443}
M.~Shifman, A.~Vainshtein and V.~Zakharov, \emph{Remarks on higgs-boson
  interactions with nucleons},
  \href{http://dx.doi.org/https://doi.org/10.1016/0370-2693(78)90481-1}{\emph{Physics
  Letters B} {\bf 78} (1978) 443 -- 446}.

\bibitem{Drees:1993bu}
M.~Drees and M.~Nojiri, \emph{{Neutralino - nucleon scattering revisited}},
  \href{http://dx.doi.org/10.1103/PhysRevD.48.3483}{\emph{Phys. Rev. D} {\bf
  48} (1993) 3483--3501}, [\href{http://arxiv.org/abs/hep-ph/9307208}{{\tt
  hep-ph/9307208}}].

\bibitem{Djouadi:2000ck}
A.~Djouadi and M.~Drees, \emph{{QCD corrections to neutralino nucleon
  scattering}},
  \href{http://dx.doi.org/10.1016/S0370-2693(00)00661-4}{\emph{Phys. Lett. B}
  {\bf 484} (2000) 183--191}, [\href{http://arxiv.org/abs/hep-ph/0004205}{{\tt
  hep-ph/0004205}}].

\bibitem{Belanger:2008sj}
G.~Belanger, F.~Boudjema, A.~Pukhov and A.~Semenov, \emph{{Dark matter direct
  detection rate in a generic model with micrOMEGAs 2.2}},
  \href{http://dx.doi.org/10.1016/j.cpc.2008.11.019}{\emph{Comput. Phys.
  Commun.} {\bf 180} (2009) 747--767},
  [\href{http://arxiv.org/abs/0803.2360}{{\tt 0803.2360}}].

\bibitem{Thomas:2012tg}
A.~Thomas, P.~Shanahan and R.~Young, \emph{{Strangeness in the nucleon: what
  have we learned?}},
  \href{http://dx.doi.org/10.1393/ncc/i2012-11292-7}{\emph{Nuovo Cim. C} {\bf
  035N04} (2012) 3--10}, [\href{http://arxiv.org/abs/1202.6407}{{\tt
  1202.6407}}].

\bibitem{Belanger:2013oya}
G.~Belanger, F.~Boudjema, A.~Pukhov and A.~Semenov, \emph{{micrOMEGAs$_3$: A
  program for calculating dark matter observables}},
  \href{http://dx.doi.org/10.1016/j.cpc.2013.10.016}{\emph{Comput. Phys.
  Commun.} {\bf 185} (2014) 960--985},
  [\href{http://arxiv.org/abs/1305.0237}{{\tt 1305.0237}}].

\bibitem{Alarcon:2011zs}
J.~Alarcon, J.~Martin~Camalich and J.~Oller, \emph{{The chiral representation
  of the $\pi N$ scattering amplitude and the pion-nucleon sigma term}},
  \href{http://dx.doi.org/10.1103/PhysRevD.85.051503}{\emph{Phys. Rev. D} {\bf
  85} (2012) 051503}, [\href{http://arxiv.org/abs/1110.3797}{{\tt 1110.3797}}].

\bibitem{Crivellin:2013ipa}
A.~Crivellin, M.~Hoferichter and M.~Procura, \emph{{Accurate evaluation of
  hadronic uncertainties in spin-independent WIMP-nucleon scattering:
  Disentangling two- and three-flavor effects}},
  \href{http://dx.doi.org/10.1103/PhysRevD.89.054021}{\emph{Phys. Rev. D} {\bf
  89} (2014) 054021}, [\href{http://arxiv.org/abs/1312.4951}{{\tt 1312.4951}}].

\bibitem{Hoferichter:2015dsa}
M.~Hoferichter, J.~Ruiz~de Elvira, B.~Kubis and U.-G. Mei\ss{}ner,
  \emph{{High-Precision Determination of the Pion-Nucleon \ensuremath{\sigma}
  Term from Roy-Steiner Equations}},
  \href{http://dx.doi.org/10.1103/PhysRevLett.115.092301}{\emph{Phys. Rev.
  Lett.} {\bf 115} (2015) 092301}, [\href{http://arxiv.org/abs/1506.04142}{{\tt
  1506.04142}}].

\bibitem{PhysRevD.101.012002}
{\scshape ATLAS} collaboration, G.~Aad et~al., \emph{Combined measurements of
  higgs boson production and decay using up to $80\text{ }\text{
  }{\mathrm{fb}}^{\ensuremath{-}1}$ of proton-proton collision data at
  $\sqrt{s}=13\text{ }\text{ }\mathrm{TeV}$ collected with the atlas
  experiment}, \href{http://dx.doi.org/10.1103/PhysRevD.101.012002}{\emph{Phys.
  Rev. D} {\bf 101} (Jan, 2020) 012002}.

\bibitem{Kang:2007ib}
J.~Kang, P.~Langacker and B.~D. Nelson, \emph{{Theory and Phenomenology of
  Exotic Isosinglet Quarks and Squarks}},
  \href{http://dx.doi.org/10.1103/PhysRevD.77.035003}{\emph{Phys. Rev. D} {\bf
  77} (2008) 035003}, [\href{http://arxiv.org/abs/0708.2701}{{\tt 0708.2701}}].

\bibitem{Babu:2008ge}
K.~Babu, I.~Gogoladze, M.~U. Rehman and Q.~Shafi, \emph{{Higgs Boson Mass,
  Sparticle Spectrum and Little Hierarchy Problem in Extended MSSM}},
  \href{http://dx.doi.org/10.1103/PhysRevD.78.055017}{\emph{Phys. Rev. D} {\bf
  78} (2008) 055017}, [\href{http://arxiv.org/abs/0807.3055}{{\tt 0807.3055}}].

\bibitem{Graham:2009gy}
P.~W. Graham, A.~Ismail, S.~Rajendran and P.~Saraswat, \emph{{A Little Solution
  to the Little Hierarchy Problem: A Vector-like Generation}},
  \href{http://dx.doi.org/10.1103/PhysRevD.81.055016}{\emph{Phys. Rev. D} {\bf
  81} (2010) 055016}, [\href{http://arxiv.org/abs/0910.3020}{{\tt 0910.3020}}].

\bibitem{Martin:2010dc}
S.~P. Martin, \emph{{Raising the Higgs Mass with Yukawa Couplings for
  Isotriplets in Vector-Like Extensions of Minimal Supersymmetry}},
  \href{http://dx.doi.org/10.1103/PhysRevD.82.055019}{\emph{Phys. Rev. D} {\bf
  82} (2010) 055019}, [\href{http://arxiv.org/abs/1006.4186}{{\tt 1006.4186}}].

\bibitem{Contino:2006qr}
R.~Contino, L.~Da~Rold and A.~Pomarol, \emph{{Light custodians in natural
  composite Higgs models}},
  \href{http://dx.doi.org/10.1103/PhysRevD.75.055014}{\emph{Phys. Rev. D} {\bf
  75} (2007) 055014}, [\href{http://arxiv.org/abs/hep-ph/0612048}{{\tt
  hep-ph/0612048}}].

\bibitem{Anastasiou:2009rv}
C.~Anastasiou, E.~Furlan and J.~Santiago, \emph{{Realistic Composite Higgs
  Models}}, \href{http://dx.doi.org/10.1103/PhysRevD.79.075003}{\emph{Phys.
  Rev. D} {\bf 79} (2009) 075003}, [\href{http://arxiv.org/abs/0901.2117}{{\tt
  0901.2117}}].

\bibitem{Vignaroli:2012sf}
N.~Vignaroli, \emph{{Discovering the composite Higgs through the decay of a
  heavy fermion}}, \href{http://dx.doi.org/10.1007/JHEP07(2012)158}{\emph{JHEP}
  {\bf 07} (2012) 158}, [\href{http://arxiv.org/abs/1204.0468}{{\tt
  1204.0468}}].

\bibitem{DeSimone:2012fs}
A.~De~Simone, O.~Matsedonskyi, R.~Rattazzi and A.~Wulzer, \emph{{A First Top
  Partner Hunter's Guide}},
  \href{http://dx.doi.org/10.1007/JHEP04(2013)004}{\emph{JHEP} {\bf 04} (2013)
  004}, [\href{http://arxiv.org/abs/1211.5663}{{\tt 1211.5663}}].

\bibitem{Agashe:2003zs}
K.~Agashe, A.~Delgado, M.~J. May and R.~Sundrum, \emph{{RS1, custodial isospin
  and precision tests}},
  \href{http://dx.doi.org/10.1088/1126-6708/2003/08/050}{\emph{JHEP} {\bf 08}
  (2003) 050}, [\href{http://arxiv.org/abs/hep-ph/0308036}{{\tt
  hep-ph/0308036}}].

\bibitem{Agashe:2004bm}
K.~Agashe and G.~Servant, \emph{{Baryon number in warped GUTs: Model building
  and (dark matter related) phenomenology}},
  \href{http://dx.doi.org/10.1088/1475-7516/2005/02/002}{\emph{JCAP} {\bf 02}
  (2005) 002}, [\href{http://arxiv.org/abs/hep-ph/0411254}{{\tt
  hep-ph/0411254}}].

\bibitem{Agashe:2004cp}
K.~Agashe, G.~Perez and A.~Soni, \emph{{Flavor structure of warped extra
  dimension models}},
  \href{http://dx.doi.org/10.1103/PhysRevD.71.016002}{\emph{Phys. Rev. D} {\bf
  71} (2005) 016002}, [\href{http://arxiv.org/abs/hep-ph/0408134}{{\tt
  hep-ph/0408134}}].

\bibitem{Contino:2008hi}
R.~Contino and G.~Servant, \emph{{Discovering the top partners at the LHC using
  same-sign dilepton final states}},
  \href{http://dx.doi.org/10.1088/1126-6708/2008/06/026}{\emph{JHEP} {\bf 06}
  (2008) 026}, [\href{http://arxiv.org/abs/0801.1679}{{\tt 0801.1679}}].

\bibitem{Gopalakrishna:2011ef}
S.~Gopalakrishna, T.~Mandal, S.~Mitra and R.~Tibrewala, \emph{{LHC Signatures
  of a Vector-like b'}},
  \href{http://dx.doi.org/10.1103/PhysRevD.84.055001}{\emph{Phys. Rev. D} {\bf
  84} (2011) 055001}, [\href{http://arxiv.org/abs/1107.4306}{{\tt 1107.4306}}].

\bibitem{Gopalakrishna:2013hua}
S.~Gopalakrishna, T.~Mandal, S.~Mitra and G.~Moreau, \emph{{LHC Signatures of
  Warped-space Vectorlike Quarks}},
  \href{http://dx.doi.org/10.1007/JHEP08(2014)079}{\emph{JHEP} {\bf 08} (2014)
  079}, [\href{http://arxiv.org/abs/1306.2656}{{\tt 1306.2656}}].

\bibitem{Han:2003wu}
T.~Han, H.~E. Logan, B.~McElrath and L.-T. Wang, \emph{{Phenomenology of the
  little Higgs model}},
  \href{http://dx.doi.org/10.1103/PhysRevD.67.095004}{\emph{Phys. Rev. D} {\bf
  67} (2003) 095004}, [\href{http://arxiv.org/abs/hep-ph/0301040}{{\tt
  hep-ph/0301040}}].

\bibitem{Carena:2006jx}
M.~Carena, J.~Hubisz, M.~Perelstein and P.~Verdier, \emph{{Collider signature
  of T-quarks}},
  \href{http://dx.doi.org/10.1103/PhysRevD.75.091701}{\emph{Phys. Rev. D} {\bf
  75} (2007) 091701}, [\href{http://arxiv.org/abs/hep-ph/0610156}{{\tt
  hep-ph/0610156}}].

\bibitem{Matsumoto:2008fq}
S.~Matsumoto, T.~Moroi and K.~Tobe, \emph{{Testing the Littlest Higgs Model
  with T-parity at the Large Hadron Collider}},
  \href{http://dx.doi.org/10.1103/PhysRevD.78.055018}{\emph{Phys. Rev. D} {\bf
  78} (2008) 055018}, [\href{http://arxiv.org/abs/0806.3837}{{\tt 0806.3837}}].

\bibitem{Berger:2012ec}
J.~Berger, J.~Hubisz and M.~Perelstein, \emph{{A Fermionic Top Partner:
  Naturalness and the LHC}},
  \href{http://dx.doi.org/10.1007/JHEP07(2012)016}{\emph{JHEP} {\bf 07} (2012)
  016}, [\href{http://arxiv.org/abs/1205.0013}{{\tt 1205.0013}}].

\bibitem{Aad:2020xfq}
{\scshape ATLAS} collaboration, G.~Aad et~al., \emph{{A search for the dimuon
  decay of the Standard Model Higgs boson with the ATLAS detector}},
  \href{http://arxiv.org/abs/2007.07830}{{\tt 2007.07830}}.

\bibitem{Sirunyan:2020two}
{\scshape CMS} collaboration, A.~M. Sirunyan et~al., \emph{{Evidence for Higgs
  boson decay to a pair of muons}},
  \href{http://arxiv.org/abs/2009.04363}{{\tt 2009.04363}}.

\bibitem{Bodwin:2013gca}
G.~T. Bodwin, F.~Petriello, S.~Stoynev and M.~Velasco, \emph{{Higgs boson
  decays to quarkonia and the $H\bar{c}c$ coupling}},
  \href{http://dx.doi.org/10.1103/PhysRevD.88.053003}{\emph{Phys. Rev. D} {\bf
  88} (2013) 053003}, [\href{http://arxiv.org/abs/1306.5770}{{\tt 1306.5770}}].

\bibitem{Delaunay:2013pja}
C.~Delaunay, T.~Golling, G.~Perez and Y.~Soreq, \emph{{Enhanced Higgs boson
  coupling to charm pairs}},
  \href{http://dx.doi.org/10.1103/PhysRevD.89.033014}{\emph{Phys. Rev. D} {\bf
  89} (2014) 033014}, [\href{http://arxiv.org/abs/1310.7029}{{\tt 1310.7029}}].

\bibitem{Kagan:2014ila}
A.~L. Kagan, G.~Perez, F.~Petriello, Y.~Soreq, S.~Stoynev and J.~Zupan,
  \emph{{Exclusive Window onto Higgs Yukawa Couplings}},
  \href{http://dx.doi.org/10.1103/PhysRevLett.114.101802}{\emph{Phys. Rev.
  Lett.} {\bf 114} (2015) 101802}, [\href{http://arxiv.org/abs/1406.1722}{{\tt
  1406.1722}}].

\bibitem{Perez:2015aoa}
G.~Perez, Y.~Soreq, E.~Stamou and K.~Tobioka, \emph{{Constraining the charm
  Yukawa and Higgs-quark coupling universality}},
  \href{http://dx.doi.org/10.1103/PhysRevD.92.033016}{\emph{Phys. Rev.} {\bf
  D92} (2015) 033016}, [\href{http://arxiv.org/abs/1503.00290}{{\tt
  1503.00290}}].

\bibitem{Perez:2015lra}
G.~Perez, Y.~Soreq, E.~Stamou and K.~Tobioka, \emph{{Prospects for measuring
  the Higgs boson coupling to light quarks}},
  \href{http://dx.doi.org/10.1103/PhysRevD.93.013001}{\emph{Phys. Rev. D} {\bf
  93} (2016) 013001}, [\href{http://arxiv.org/abs/1505.06689}{{\tt
  1505.06689}}].

\bibitem{Koenig:2015pha}
M.~K\"onig and M.~Neubert, \emph{{Exclusive Radiative Higgs Decays as Probes of
  Light-Quark Yukawa Couplings}},
  \href{http://dx.doi.org/10.1007/JHEP08(2015)012}{\emph{JHEP} {\bf 08} (2015)
  012}, [\href{http://arxiv.org/abs/1505.03870}{{\tt 1505.03870}}].

\bibitem{Brivio:2015fxa}
I.~Brivio, F.~Goertz and G.~Isidori, \emph{{Probing the Charm Quark Yukawa
  Coupling in Higgs+Charm Production}},
  \href{http://dx.doi.org/10.1103/PhysRevLett.115.211801}{\emph{Phys. Rev.
  Lett.} {\bf 115} (2015) 211801}, [\href{http://arxiv.org/abs/1507.02916}{{\tt
  1507.02916}}].

\bibitem{Soreq:2016rae}
Y.~Soreq, H.~X. Zhu and J.~Zupan, \emph{{Light quark Yukawa couplings from
  Higgs kinematics}},
  \href{http://dx.doi.org/10.1007/JHEP12(2016)045}{\emph{JHEP} {\bf 12} (2016)
  045}, [\href{http://arxiv.org/abs/1606.09621}{{\tt 1606.09621}}].

\bibitem{Yu:2016rvv}
F.~Yu, \emph{{Phenomenology of Enhanced Light Quark Yukawa Couplings and the
  $W^\pm h$ Charge Asymmetry}},
  \href{http://dx.doi.org/10.1007/JHEP02(2017)083}{\emph{JHEP} {\bf 02} (2017)
  083}, [\href{http://arxiv.org/abs/1609.06592}{{\tt 1609.06592}}].

\bibitem{Cohen:2017rsk}
J.~Cohen, S.~Bar-Shalom, G.~Eilam and A.~Soni, \emph{{Light-quarks Yukawa
  couplings and new physics in exclusive high- $p_T$ Higgs boson+jet and Higgs
  boson + b -jet events}},
  \href{http://dx.doi.org/10.1103/PhysRevD.97.055014}{\emph{Phys. Rev. D} {\bf
  97} (2018) 055014}, [\href{http://arxiv.org/abs/1705.09295}{{\tt
  1705.09295}}].

\bibitem{Han:2018juw}
T.~Han, B.~Nachman and X.~Wang, \emph{{Charm-quark Yukawa Coupling in
  $h\rightarrow c\bar{c}\gamma$ at LHC}},
  \href{http://dx.doi.org/10.1016/j.physletb.2019.04.031}{\emph{Phys. Lett. B}
  {\bf 793} (2019) 90--96}, [\href{http://arxiv.org/abs/1812.06992}{{\tt
  1812.06992}}].

\bibitem{Mao:2019hgg}
S.~Mao, Y.~Guo-He, L.~Gang, Z.~Yu and G.~Jian-You, \emph{{Probing the
  charm-Higgs Yukawa coupling via Higgs boson decay to $h_c$ plus a photon}},
  \href{http://dx.doi.org/10.1088/1361-6471/ab1a6e}{\emph{J. Phys. G} {\bf 46}
  (2019) 105008}, [\href{http://arxiv.org/abs/1905.01589}{{\tt 1905.01589}}].

\bibitem{Coyle:2019hvs}
N.~M. Coyle, C.~E. Wagner and V.~Wei, \emph{{Bounding the charm Yukawa
  coupling}}, \href{http://dx.doi.org/10.1103/PhysRevD.100.073013}{\emph{Phys.
  Rev. D} {\bf 100} (2019) 073013},
  [\href{http://arxiv.org/abs/1905.09360}{{\tt 1905.09360}}].

\bibitem{Alasfar:2019pmn}
L.~Alasfar, R.~Corral~Lopez and R.~Gröber, \emph{{Probing Higgs couplings to
  light quarks via Higgs pair production}},
  \href{http://dx.doi.org/10.1007/JHEP11(2019)088}{\emph{JHEP} {\bf 11} (2019)
  088}, [\href{http://arxiv.org/abs/1909.05279}{{\tt 1909.05279}}].

\bibitem{Aguilar-Saavedra:2020rgo}
J.~Aguilar-Saavedra, J.~Cano and J.~No, \emph{{More light on Higgs flavor at
  the LHC: Higgs couplings to light quarks through $h + \gamma$ production}},
  \href{http://arxiv.org/abs/2008.12538}{{\tt 2008.12538}}.

\bibitem{Aad:2015sda}
{\scshape ATLAS} collaboration, G.~Aad et~al., \emph{{Search for Higgs and Z
  Boson Decays to $J/\psi \gamma and \Upsilon{}(nS)\gamma$ with the ATLAS
  Detector}},
  \href{http://dx.doi.org/10.1103/PhysRevLett.114.121801}{\emph{Phys. Rev.
  Lett.} {\bf 114} (2015) 121801}, [\href{http://arxiv.org/abs/1501.03276}{{\tt
  1501.03276}}].

\bibitem{Aaboud:2016rug}
{\scshape ATLAS} collaboration, M.~Aaboud et~al., \emph{{Search for Higgs and
  $Z$ Boson Decays to $\phi\,\gamma$ with the ATLAS Detector}},
  \href{http://dx.doi.org/10.1103/PhysRevLett.117.111802}{\emph{Phys. Rev.
  Lett.} {\bf 117} (2016) 111802}, [\href{http://arxiv.org/abs/1607.03400}{{\tt
  1607.03400}}].

\bibitem{LHCb:2016yxg}
{\scshape LHCb} collaboration, \emph{{Search for $H^0 \rightarrow b \bar{b}$ or
  $c \bar{c}$ in association with a $W$ or $Z$ boson in the forward region of
  $pp$ collisions}},  Tech. Rep. LHCb-CONF-2016-006, CERN-LHCb-CONF-2016-006,
  9, 2016.

\bibitem{Aaboud:2017xnb}
{\scshape ATLAS} collaboration, M.~Aaboud et~al., \emph{{Search for exclusive
  Higgs and $Z$ boson decays to $\phi\gamma$ and $\rho\gamma$ with the ATLAS
  detector}}, \href{http://dx.doi.org/10.1007/JHEP07(2018)127}{\emph{JHEP} {\bf
  07} (2018) 127}, [\href{http://arxiv.org/abs/1712.02758}{{\tt 1712.02758}}].

\bibitem{Aaboud:2018fhh}
{\scshape ATLAS} collaboration, M.~Aaboud et~al., \emph{{Search for the Decay
  of the Higgs Boson to Charm Quarks with the ATLAS Experiment}},
  \href{http://dx.doi.org/10.1103/PhysRevLett.120.211802}{\emph{Phys. Rev.
  Lett.} {\bf 120} (2018) 211802}, [\href{http://arxiv.org/abs/1802.04329}{{\tt
  1802.04329}}].

\bibitem{Sirunyan:2020mds}
{\scshape CMS} collaboration, A.~M. Sirunyan et~al., \emph{{Search for decays
  of the 125 GeV Higgs boson into a Z boson and a $\rho$ or $\phi$ meson}},
  \href{http://arxiv.org/abs/2007.05122}{{\tt 2007.05122}}.

\bibitem{Goertz:2014qia}
F.~Goertz, \emph{{Indirect Handle on the Down-Quark Yukawa Coupling}},
  \href{http://dx.doi.org/10.1103/PhysRevLett.113.261803}{\emph{Phys. Rev.
  Lett.} {\bf 113} (2014) 261803}, [\href{http://arxiv.org/abs/1406.0102}{{\tt
  1406.0102}}].

\bibitem{Delaunay:2016brc}
C.~Delaunay, R.~Ozeri, G.~Perez and Y.~Soreq, \emph{{Probing Atomic Higgs-like
  Forces at the Precision Frontier}},
  \href{http://dx.doi.org/10.1103/PhysRevD.96.093001}{\emph{Phys. Rev. D} {\bf
  96} (2017) 093001}, [\href{http://arxiv.org/abs/1601.05087}{{\tt
  1601.05087}}].

\bibitem{Gao:2016jcm}
J.~Gao, \emph{{Probing light-quark Yukawa couplings via hadronic event shapes
  at lepton colliders}},
  \href{http://dx.doi.org/10.1007/JHEP01(2018)038}{\emph{JHEP} {\bf 01} (2018)
  038}, [\href{http://arxiv.org/abs/1608.01746}{{\tt 1608.01746}}].

\bibitem{Belanger:2006is}
G.~Belanger, F.~Boudjema, A.~Pukhov and A.~Semenov, \emph{{MicrOMEGAs 2.0: A
  Program to calculate the relic density of dark matter in a generic model}},
  \href{http://dx.doi.org/10.1016/j.cpc.2006.11.008}{\emph{Comput. Phys.
  Commun.} {\bf 176} (2007) 367--382},
  [\href{http://arxiv.org/abs/hep-ph/0607059}{{\tt hep-ph/0607059}}].

\bibitem{Jarosik_2011}
N.~Jarosik, C.~L. Bennett, J.~Dunkley, B.~Gold, M.~R. Greason, M.~Halpern
  et~al., \emph{Seven-year wilkinson microwave anisotropy probe ( wmap )
  observations: Sky maps, systematic errors, and basic results},
  \href{http://dx.doi.org/10.1088/0067-0049/192/2/14}{\emph{The Astrophysical
  Journal Supplement Series} {\bf 192} (Jan, 2011) 14}.

\bibitem{Aghanim:2018eyx}
{\scshape Planck} collaboration, N.~Aghanim et~al., \emph{{Planck 2018 results.
  VI. Cosmological parameters}},  \href{http://arxiv.org/abs/1807.06209}{{\tt
  1807.06209}}.

\bibitem{Bhattacharya:2017fid}
S.~Bhattacharya, P.~Ghosh, T.~N. Maity and T.~S. Ray, \emph{{Mitigating Direct
  Detection Bounds in Non-minimal Higgs Portal Scalar Dark Matter Models}},
  \href{http://dx.doi.org/10.1007/JHEP10(2017)088}{\emph{JHEP} {\bf 10} (2017)
  088}, [\href{http://arxiv.org/abs/1706.04699}{{\tt 1706.04699}}].

\bibitem{Yepes:2018zkk}
J.~Yepes, \emph{{Top partners tackling vector dark matter}},
  \href{http://dx.doi.org/10.1016/j.physletb.2020.135890}{\emph{Phys. Lett. B}
  {\bf 811} (2020) 135890}, [\href{http://arxiv.org/abs/1811.06059}{{\tt
  1811.06059}}].

\bibitem{Korber:2017ery}
C.~K\"orber, A.~Nogga and J.~de~Vries, \emph{{First-principle calculations of
  Dark Matter scattering off light nuclei}},
  \href{http://dx.doi.org/10.1103/PhysRevC.96.035805}{\emph{Phys. Rev. C} {\bf
  96} (2017) 035805}, [\href{http://arxiv.org/abs/1704.01150}{{\tt
  1704.01150}}].

\end{thebibliography}\endgroup
\bibliographystyle{JHEPCust}
\end{document}